\documentclass[a4paper,fleqn,usenatbib]{mnras}

\usepackage{newtxtext,newtxmath}

\usepackage[T1]{fontenc}
\usepackage{ae,aecompl}

\usepackage{graphicx}	
\usepackage{amsmath}	
\usepackage{amssymb}	
\usepackage{epsfig}

\title[Evolution of substructures in clusters]{Evolution of the degree of substructures in simulated
galaxy clusters}

\author[C. De Boni et al.]{
Cristiano De Boni$^{1}$\thanks{E-mail: cdeboni@mpe.mpg.de (CDB)},
Hans B\"ohringer$^{1}$,
Gayoung Chon$^{1}$
and Klaus Dolag$^{2,3}$
\\
$^{1}$Max-Planck-Institut f\"ur extraterrestrische Physik (MPE), Gie{\ss}enbachstra{\ss}e, D-85741 Garching bei M\"unchen, Germany \\
$^2$University Observatory Munich, Scheinerstr. 1, D-81679 Munich, Germany \\
$^3$Max-Planck-Institut f\"ur Astrophysik (MPA), Karl-Schwarzschild Strasse 1, D-85748 Garching bei M\"unchen, Germany \\
}

\date{Accepted XXX. Received YYY; in original form ZZZ}

\pubyear{2018}

\begin{document}
\label{firstpage}
\pagerange{\pageref{firstpage}--\pageref{lastpage}}
\maketitle

\begin{abstract}
We study the evolution of substructure in the mass distribution with mass, redshift and radius in a sample of simulated galaxy clusters. The sample, containing $1226$ objects, spans the mass range $M_{200} = 10^{14} - 1.74 \times 10^{15} \ {\rm M_{\odot}} \ h^{-1}$ in six redshift bins from $z=0$ to $z=1.179$. We consider three different diagnostics: 1) subhalos identified with SUBFIND; 2) overdense regions localized by dividing the cluster into octants; 3) offset between the potential minimum and the center of mass. The octant analysis is a new method that we introduce in this work. We find that none of the diagnostics indicate a correlation between the mass of the cluster and the fraction of substructures. On the other hand, all the diagnostics suggest an evolution of substructures with redshift. For SUBFIND halos, the mass fraction is constant with redshift at $R_{\mathrm{vir}}$, but shows a mild evolution at $R_{200}$ and $R_{500}$. Also, the fraction of clusters with at least a subhalo more massive than one thirtieth of the total mass is less than $20 \%$. Our new method based on the octants returns a mass fraction in substructures which has a strong evolution with redshift at all radii. The offsets also evolve strongly with redshift. We also find a strong correlation for individual clusters between the offset and the fraction of substructures identified with the octant analysis. Our work puts strong constraints on the amount of substructures we expect to find in galaxy clusters and on their evolution with redshift.
\end{abstract}

\begin{keywords}
galaxies: clusters: general -- methods: numerical 
\end{keywords}



\section{Introduction}

Galaxy clusters are an important building block of modern cosmology. Cluster number counts and mass function are used as cosmological probe to constrain cosmological parameters, alone or in combination with other kind of data \citep[{\it e.g.} \rm][and references therein]{2016A&A...594A..24P,2014A&A...570A..31B}. Considerable effort has lead nowadays to the discovery of thousands of clusters through optical \citep{2014ApJ...785..104R, 2016ApJS..224....1R}, X-ray \citep{2004A&A...425..367B} and Sunyaev Zel'dovich (SZ) effect \citep{2016A&A...594A..27P,2011A&A...536A...8P,2013JCAP...07..008H,2015ApJS..216...27B} observations. 

To determine the masses of a large sample of clusters for cosmological studies, one relies on scaling relations between mass and observed quantities \citep[e.g.][]{2009A&A...498..361P,2016MNRAS.463.3582M,2017MNRAS.468.3347S} which are calibrated on a smaller sample of clusters. Scaling relations evolve with redshift, but it is still unclear whether this happens because of a different substructure fraction or because relaxed clusters themselves evolve. This will also affect the concentration-mass relation. As it has been pointed out in several works that there is a difference in the scaling relations for regular and disturbed clusters \cite[{\it e.g.} \rm][]{2012A&A...548A..59C}, it is important to take this effect into account for cosmological and astrophysical modeling based on scaling relations. Different cluster samples may have different fractions of regular and disturbed clusters, leading to different effects on scaling relations. \cite{2016A&A...585A.147A} pointed out that a dynamical mass selected sample shows an higher scatter and a different slope scaling relation with respect to an X-ray selected sample. \cite{2017A&A...606L...4C} showed that a volume-limited sample shows an higher fraction of dynamically disturbed clusters with respect to a flux-limited one. Since the morphology of clusters is also important in discussing the cool-core distribution \citep{2010A&A...514A..32B,2017MNRAS.468.1917R}, it is evident that a proper treatment of the dynamical state of a sample is mandatory from an observational perspective. In this context it is important to know to which degree the morphological fractions are dependent on mass and redshift evolution.

In this paper we study the evolution of the fraction of substructures in a sample of simulated galaxy clusters, as a function of mass, radius and redshift. We test different methods to determine and compare the substructure distribution and the fraction of mass associated to these substructures. In particular, in addition to discussing two definitions of substructures traditionally used in the literature like SUBFIND and the offset between the potential minimum and the center of mass, we present a new method based on the division of clusters into octants. By searching for localized matter overdensities in each octant, we determine an excess in mass that we associate to the presence of substructures. We want to understand and characterize the amount of substructures and their evolution with redshift.

The classification between relaxed and unrelaxed cluster is important also in theoretical and numerical works, in particular in the study of the relation between concentration and mass \citep[{\it e.g.} \rm][]{2013MNRAS.428.2921D}. At a fixed redshift, more massive clusters are expected to be younger and thus to have a more disturbed dynamical state with respect to less massive objects. Different authors use different criteria to determine the dynamical state of clusters in numerical simulations. A widely used quantity is the offset between the potential minimum of the cluster and the center of mass of the matter distribution within a given radius. Most of the authors quote an offset of $0.07$ of the reference radius as the maximum value for being considered relaxed \citep{2007MNRAS.381.1450N,2008MNRAS.391.1940M,2008MNRAS.390L..64D,2016MNRAS.460.1214L}. \cite{2008MNRAS.391.1940M} use the residual of the best-fit of a NFW profile to quantify the dynamical state. \cite{2007MNRAS.381.1450N} and \cite{2016MNRAS.460.1214L} use the so called virial ratio, {\it i.e.} the ratio between twice the kinetic energy and the absolute value of the potential energy, with a maximum threshold of $1.35$. The same authors also use the fraction of mass in substructures determined with SUBFIND as an indicator for the dynamical state, and set the maximum value allowed for a relaxed cluster to $0.1$, meaning that cluster having more than $10 \%$ of their mass in substructures are considered to be not relaxed.

The paper is organized as follows: in Section \ref{simulations} we describe the numerical simulations we use for this analysis and in Section \ref{sample} we present our selected sample. The methods to investigate substructures in clusters are presented in Section \ref{methods} while the evolution of substructures with mass and redshift is discussed in Section \ref{evolution}. We draw our conclusions in Section \ref{conclusions}.

\section{Magneticum simulations} \label{simulations}

The Magneticum Simulations are a set of cosmological hydrodynamical simulations with different box size, ranging from $2688$ to $18 \ {\rm Mpc} \ h^{-1}$, and mass resolution, ranging from $1.3 \times 10^{10}$ to $1.9 \times 10^6 \ {\rm M_{\odot}} \ h^{-1}$, run with the GADGET code \cite{2005MNRAS.364.1105S}. The simulations are normalized to WMAP7 cosmology \cite{2011ApJS..192...18K}, namely total matter density $\Omega_{m,0}=0.272$ (with baryons counting for $16.8 \%$), cosmological constant $\Omega_{\Lambda,0}=0.728$, Hubble constant $H_0=70.4$, spectral index of the primordial power spectrum $n_{s}=0.963$ and normalization of the power spectrum $\sigma_{8}=0.809$. These simulations include cooling, star formation, winds \citep{2003MNRAS.339..289S}, metals, stellar population and chemical enrichment from SN-Ia, SN-II and AGB \citep{2003MNRAS.342.1025T,2007MNRAS.382.1050T}, black holes and AGN feedback \citep{2005MNRAS.361..776S,2010MNRAS.401.1670F,2014MNRAS.442.2304H}, thermal conduction \citep{2004ApJ...606L..97D} and magnetic fields \citep{2009MNRAS.398.1678D}.

For this study, we use the so-called Box1 of the Magneticum Simulation, with medium mass resolution. The box size is $896 \ {\rm Mpc} \ h^{-1}$, populated with $2 \times 1526^3$ particles, half of which are dark matter particles, while the other half is constituted by (hot and cold) gas particles. The mass resolution is $1.3 \times 10^{10} \ {\rm M_{\odot}} \ h^{-1}$ for dark matter particles and $2.6 \times 10^9 \ {\rm M_{\odot}} \ h^{-1}$ for gas particles. The softening length is $10 \ {\rm kpc} \ h^{-1}$ for dark matter and gas particles, while it is $5 \ {\rm kpc} \ h^{-1}$ for star particles.

\subsection{Halo definition}

Halos in Magneticum Simulations are found, on the fly, through a friends-of-friends (FoF) algorithm. Particles in the simulation are linked together according to their distance if this falls below a given linking length. A linking length of $0.16$ times the average interparticle separation was used in this run. This linking defines ``FoF halos'' within the cosmological box. The center of a FoF halo is assigned to the position of the most bound particle in the halo, {\it i.e.} to the location of the minimum of the gravitational potential.

Starting from the center of a halo, spherical overdensity cluster masses can be evaluated as the mass within a sphere enclosing a density which is a given value $\Delta$ (dubbed overdensity) of the critical density of the Universe at the redshift of the halo. The radius and the corresponding enclosed mass are indicated as $R_{\Delta}$ and $M_{\Delta}$. Typical values for $\Delta$, both in theoretical and observational studies, are $2500$, $500$ and $200$. 
Another value which is often considered comes from the spherical top-hat collapse model, parametrized by the \cite{1998ApJ...495...80B} formula:

\begin{equation}
\Delta_{th} = 18 \pi^2 + 82 x - 39 x^2 \ ,
\end{equation}

\noindent where $x= \left[ \frac{\Omega_{m,0}(1+z)}{E_z^2} -1 \right]$ and $E_z = H(z) / H_0$. This value is commonly used to define the virial overdensity, so in the following we will use $R_{\mathrm{vir}}$ and $M_{\mathrm{vir}}$ to indicate the corresponding radius and mass.
This overdensity evolves with time taking into account the background evolution of the Universe. Nevertheless, like for the fixed overdensities, the enclosed mass suffers from the so-called ``pseudo-evolution'', as pointed out by \cite{2013ApJ...766...25D}.

\subsection{SUBFIND subhalos}

The SUBFIND algorithm of the GADGET code divides every FoF halo into subhalos according to a density threshold combined with a self-boundedness requirement, as described in detail in \cite{2001MNRAS.328..726S}. If, after the unbinding, a minimum number of $20$ particles survives, this set of particles constitutes a subhalo of the original FoF halo. We refer to the most massive subhalo identified by SUBFIND as ``main halo'' and to the other subhalos as ``satellite'' halos.
In summary, SUBFIND identifies self-bound structures within the density field of the original structure identified by the FoF algorithm. This process is run on the fly by the simulation while identifying the FoF halos.

\section{Cluster selection} \label{sample}

From the Magneticum Simulations, we select a sample of massive clusters at different redshifts. The selection is performed randomly based only on cluster mass, without any {\it a priori} knowledge of the dynamical state of the cluster or of the X-ray properties. With the REFLEX nearby sample in mind for a comparison \citep{2017A&A...606L...4C}, we perform our reference selection at $z=0.137$ and propagate this selection to other redshift. We select our clusters from six snapshots of the simulation, spanning the redshift range $z=0-1.179$. We consider all the objects with $M_{200} \geq 10^{14} \ {\rm M_{\odot}} \ h^{-1}$ for which the center is not closer than $5 \ {\rm Mpc} \ h^{-1}$ to the border of the box, thus reducing the effective box size to $886 \ {\rm Mpc} \ h^{-1}$. We build $10$ mass bins containing $25$ objects each. We construct the first mass bin by taking the $25$ most massive clusters in the box at $z=0.137$. Then, we divide the interval from $10^{14} \ {\rm M_{\odot}} \ h^{-1}$ to the mass of the $25$-th most massive object in $9$ logarithmic spaced mass bins. Finally, we randomly select $25$ objects in each mass bin. So, at the reference redshift, we end up with $250$ clusters. We keep the mass bins defined in this way also for randomly selecting the clusters at other redshifts. This is reflected in a decrease of the number of halos in the high-mass bins at higher redshift, as a consequence of the evolution of cluster mass function. 
The total number of clusters that end up in the different mass bins at the various redshifts are summarized in Table \ref{sample_table}. Our final sample consists of 1226 clusters.

\begin{table*}
\begin{center}
\caption{Mass bins, selected sample and total number of clusters for the six redshifts considered. The fractions in the redshift columns indicate the ratio between the number of selected clusters and the total number of clusters in the mass bins.}  \label{sample_table}
\begin{tabular}{cccccccccc}
\\
\hline
\hline
Bin & $M_{\mathrm{min}}$ & $M_{\mathrm{max}}$ & [$10^{14} \ {\rm M_{\odot}} \ h^{-1}$] & $z=0$ & $z=0.137$ & $z=0.293$ & $z=0.470$ & $z=0.782$ & $z=1.179$ \\
\hline
$10$ & $9.116$ & $17.379$ & $$ & $25/47$ & $25/25$ &$12/12$ & $2/2$ & $-$ & $-$ \\
\hline
$9$ & $7.131$& $9.116$ & $$ & $25/47$ & $25/40$ & $21/21$ & $11/11$ & $-$ & $-$ \\
\hline
$8$ & $5.578$ & $7.131$ & $$ & $25/106$ & $25/63$ & $25/42$ & $22/22$ & $6/6$ & $-$ \\
\hline
$7$ & $4.364$ & $5.578$ & $$ & $25/188$ & $25/155$ & $25/94$ & $25/58$ & $10/10$ & $1/1$ \\
\hline
$6$ & $3.413$ & $4.364$ & $$ & $25/352$ & $25/285$ & $25/197$ & $25/104$ & $25/29$ & $6/6$ \\
\hline
$5$ & $2.670$ & $3.413$ & $$ & $25/621$ & $25/490$ & $25/362$ & $25/219$ & $25/79$ & $10/10$ \\
\hline
$4$ & $2.089$ & $2.670$ & $$ & $25/993$ & $25/863$ & $25/622$ & $25/436$ & $25/164$ & $25/40$ \\
\hline
$3$ & $1.634$ & $2.089$ & $$ & $25/1431$ & $25/1137$ &  $25/958$ & $25/673$ & $25/309$ & $25/59$ \\
\hline
$2$ & $1.278$ & $1.634$ & $$ & $25/2106$ & $25/1828$ & $25/1502$ & $25/1160$ & $25/516$ & $25/162$ \\
\hline
$1$ & $1.000$ & $1.278$ & $$ & $25/3143$ & $25/2765$ & $25/2336$ & $25/1775$ & $25/959$ & $25/303$ \\
\hline
\hline
Total & $1.000$ & $17.379$ & $$ & $250/9304$ & $250/7651$ & $233/6146$ & $210/4460$ & $166/2072$ & $117/581$ \\
\hline
\hline
\end{tabular}
\end{center}
\end{table*}

\noindent For all the clusters in our sample we evaluate the density and mass profiles in linearly-spaced bins of $50 \ {\rm kpc} \ h^{-1}$ up to $10 \ {\rm Mpc} \ h^{-1}$ from the cluster center, which is defined as the minimum of the gravitational potential and corresponds to the location of the most bound particle. We choose this size of the radial binning because it is compatible with the minimum size of the substructures we want to characterize. For every cluster, we evaluate the spherical overdensity radii $R_{\mathrm{vir}}$, $R_{200}$ and $R_{500}$ and the corresponding enclosed masses, $M_{\mathrm{vir}}$, $M_{200}$ and $M_{500}$, respectively.

\section{Methods} \label{methods}

In this section we present three different methods to assess the substructure distribution in the cluster sample. For all methods, in general we normalize quantities for the reference radius and the enclosed mass, in order to check the self-similarity of the substructure distribution at the different radii.

\subsection{Substructures detected with SUBFIND}

We start with the study of substructures in clusters as defined by SUBFIND, for different mass and radial thresholds. We generically indicate the mass fraction in substructures identified by SUBFIND with $f_{\mathrm{sub}}$.
We are interested in the distribution of substructures within the virial radius, $R_{\mathrm{vir}}$, and two smaller radii, $R_{200}$ and $R_{500}$. For every cluster, we define the quantity $f_{\mathrm{sub,vir}}$ ($f_{\mathrm{sub},200}$, $f_{\mathrm{sub},500}$) as the ratio between the sum of the mass of all the satellite halos identified by SUBFIND within $R_{\mathrm{vir}}$ ($R_{200}$, $R_{500}$) and $M_{\mathrm{vir}}$ ($M_{200}$, $M_{500}$). In this case, we do not set a minimum mass threshold for substructures and we take all identified halos with a minimum of $20$ particles. We postpone to the Appendix \ref{outskirts} the discussion of the substructure distribution outside the virial radius. 

\begin{figure}
\hbox{
  \epsfig{figure=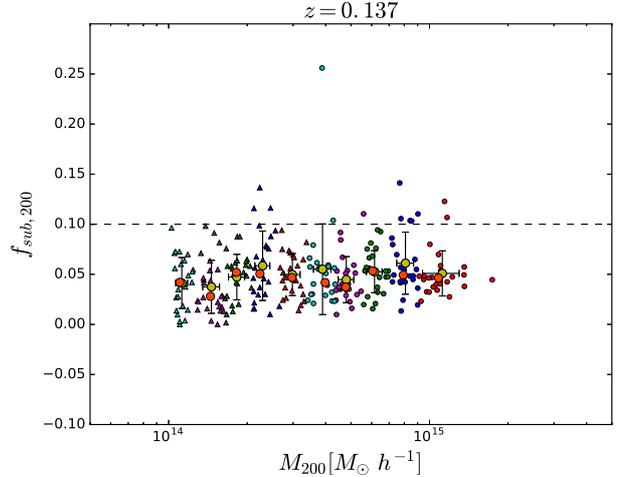,width=0.5\textwidth}
}
\caption{Distribution of $f_{\mathrm{sub},200}$ at $z=0.137$. Circles mark the mass bins from 10 to 6, triangles mark the mass bins form 5 to 1. The big yellow circles mark the mean values in every mass bin, the big orange circles mark the median. Error bars mark one standard deviation of the mean. The horizontal dashed-line indicates $f_{\mathrm{sub},200} = 0.1$, which is commonly used in the literature as a threshold to divide relaxed and unrelaxed clusters.}
\label{f_sub_200_figure}
\end{figure}

Figure \ref{f_sub_200_figure} shows an example of the distribution of $f_{\mathrm{sub},200}$ at our reference redshift $z=0.137$. There is no strong correlation between $f_{\mathrm{sub},200}$ and $M_{200}$ and there is a large scatter at fixed mass. This is the case also for the other redshifts as well as for the other radii. The mean values are strongly influenced by a small number of clusters with a high mass fraction in substructures, as it is evident by comparing the mean with the median, which is always lower. The fact that there is no obvious dependence on mass means that we can consider all the clusters in a given snapshot independently on their mass, and evaluate average properties at every redshift. We report the results for $f_{\mathrm{sub,vir}}$, $f_{\mathrm{sub},200}$ and $f_{\mathrm{sub},500}$ in Table \ref{f_sub_table}. In this table we quote the mean and the standard deviation, along with the median, the $16 \%$ and $84 \%$ percentiles.

\begin{table*}
\begin{center}
\caption{Distribution of $f_{\mathrm{sub,vir}}$, $f_{\mathrm{sub},200}$ and $f_{\mathrm{sub},500}$ at different redshifts. The mean and median are evaluated over all the clusters at a given redshift. We also quote the standard error, and the $16 \%$ and $84 \%$ percentiles of the distribution.}  \label{f_sub_table}
\begin{tabular}{cccccccccccccccc} 
\\
\hline
\hline
 & \multicolumn{5}{c}{$f_{\mathrm{sub,vir}}$} & \multicolumn{5}{c}{$f_{\mathrm{sub},200}$} & \multicolumn{5}{c}{$f_{\mathrm{sub},500}$} \\
\hline
Redshift & mean & $\sigma$ & median & $16 \%$ & $84 \%$ & \ mean & $\sigma$ & median & $16 \%$ & $84 \%$ & \ mean & $\sigma$ & median & $16 \%$ & $84 \%$ \\
\hline
$z=0$ & $0.057$ & $0.029$ & $0.055$ & $0.030$ & $0.082$ & $\ 0.049$ & $0.028$ & $0.048$ & $0.021$ & $\ 0.074$ & $0.036$ & $0.030$ & $0.028$ & $0.009$ & $0.062$ \\
\hline
$z=0.137$ & $0.057$ & $0.033$ & $0.051$ & $0.031$ & $0.080$ & $\ 0.050$ & $0.029$ & $0.047$ & $0.023$ & $0.074$ & $\ 0.037$ & $0.026$ & $0.030$ & $0.015$ & $0.062$ \\
\hline
$z=0.293$ & $0.058$ & $0.031$ & $0.053$ & $0.030$ & $0.082$ & $\ 0.052$ & $0.029$ & $0.050$ & $0.028$ & $0.075$ & $\ 0.040$ & $0.028$ & $0.035$ & $0.017$ & $0.061$ \\
\hline
$z=0.470$ & $0.060$ & $0.036$ & $0.053$ & $0.032$ & $0.085$ & $\ 0.055$ & $0.030$ & $0.050$ & $0.029$ & $0.077$ & $\ 0.043$ & $0.027$ & $0.042$ & $0.016$ & $0.066$ \\
\hline
$z=0.783$ & $0.057$ & $0.033$ & $0.050$ & $0.029$ & $0.082$ & $\ 0.056$ & $0.033$ & $0.049$ & $0.029$ & $0.084$ & $\ 0.044$ & $0.032$ & $0.038$ & $0.016$ & $0.071$ \\
\hline
$z=1.179$ & $0.056$ & $0.036$ & $0.047$  & $0.026$ & $0.087$ & $\ 0.053$ & $0.031$ & $0.046$ & $0.025$ & $0.076$ & $\ 0.047$ & $0.029$ & $0.046$ & $0.020$ & $0.063$ \\
\hline
\hline
\end{tabular}
\end{center}
\end{table*}

To distinguish the effect of galaxy-size halos and larger substructures, we are also interested in checking the fraction in substructures more massive than a given threshold. We set this threshold at $10^{12} \ {\rm M_{\odot}} \ h^{-1}$, to exclude galaxy-sized halos from the analysis, and call the corresponding fraction $f_{\mathrm{sub,vir},>1e12}$ ($f_{\mathrm{sub},200,>1e12}$, $f_{\mathrm{sub},500,>1e12}$). We then evaluate the ratio $f_{\mathrm{sub,vir},>1e12}/f_{\mathrm{sub,vir}}$ ($f_{\mathrm{sub},200,>1e12}/f_{\mathrm{sub},200}$, $f_{\mathrm{sub},500,>1e12}/f_{\mathrm{sub},500}$) individually for every cluster. Also for this quantity there is no strong correlation with mass. We report the results in Table \ref{f_sub_mass_ratio_table}.

\begin{table*}
\begin{center}
\caption{Same as Table \ref{f_sub_table}, but for the individual cluster ratios $f_{\mathrm{sub,vir},>1e12}/f_{\mathrm{sub,vir}}$, $f_{\mathrm{sub},200,>1e12}/f_{\mathrm{sub},200}$ and $f_{\mathrm{sub},500,>1e12}/f_{\mathrm{sub},500}$.}  \label{f_sub_mass_ratio_table}
\begin{tabular}{cccccccccccccccc} 
\\
\hline
\hline
 & \multicolumn{5}{c}{$f_{\mathrm{sub,vir},>1e12}/f_{\mathrm{sub,vir}}$} & \multicolumn{5}{c}{$f_{\mathrm{sub},200,>1e12}/f_{\mathrm{sub},200}$} & \multicolumn{5}{c}{$f_{\mathrm{sub},500,>1e12}/f_{\mathrm{sub},500}$} \\
\hline
Redshift & mean & $\sigma$ & median & $16 \%$ & $84 \%$ & \ mean & $\sigma$ & median & $16 \%$ & $84 \%$ & \ mean & $\sigma$ & median & $16 \%$ & $84 \%$ \\
\hline
$z=0$ & $0.718$ & $0.201$ & $0.779$ & $0.577$ & $0.873$ & $\ 0.694$ & $0.238$ & $0.765$ & $0.516$ & $0.885$ & $\ 0.602$ & $0.332$ & $0.719$ & $0.000$ & $0.900$ \\
\hline
$z=0.137$ & $0.727$ & $0.179$ & $0.767$ & $0.585$ & $0.868$ & $\ 0.701$ & $0.218$ & $0.752$ & $0.570$ & $0.872$ & $\ 0.625$ & $0.310$ & $0.727$ & $0.252$ & $0.894$ \\
\hline
$z=0.293$ & $0.711$ & $0.191$ & $0.744$ & $0.590$ & $0.865$ & $\ 0.691$ & $0.217$ & $0.751$ & $0.553$ & $0.858$ & $\ 0.636$ & $0.294$ & $0.729$ & $0.376$ & $0.894$ \\
\hline
$z=0.470$ & $0.725$ & $0.183$ & $0.754$ & $0.615$ & $0.877$ & $\ 0.711$ & $0.197$ & $0.755$ & $0.556$ & $0.868$ & $\ 0.632$ & $0.315$ & $0.747$ & $0.093$ & $0.885$ \\
\hline
$z=0.783$ & $0.732$ & $0.187$ & $0.780$ & $0.561$ & $0.899$ & $\ 0.724$ & $0.206$ & $0.775$ & $0.561$ & $0.902$ & $\ 0.647$ & $0.314$ & $0.764$ & $0.288$ & $0.904$ \\
\hline
$z=1.179$ & $0.716$ & $0.217$ & $0.767$ & $0.535$ & $0.916$ & $\ 0.695$ & $0.234$ & $0.736$ & $0.496$ & $0.910$ & $\ 0.665$ & $0.306$ & $0.777$ & $0.406$ & $0.920$ \\
\hline
\hline
\end{tabular}
\end{center}
\end{table*}

\begin{figure}
\hbox{
  \epsfig{figure=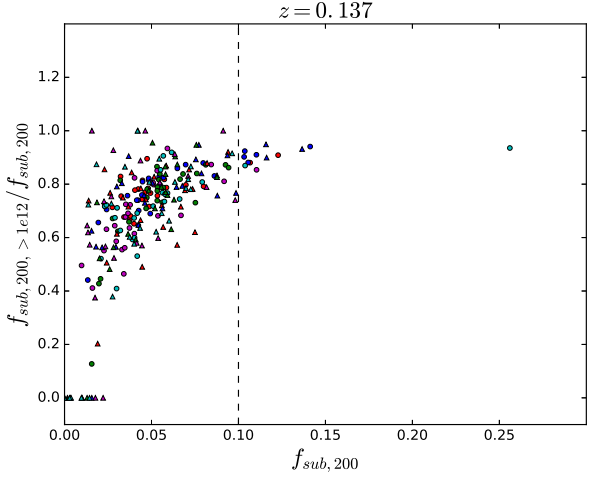,width=0.5\textwidth}
}
\caption{Individual cluster relation between $f_{\mathrm{sub},200,>1e12}/f_{\mathrm{sub},200}$ and $f_{\mathrm{sub},200}$ at $z=0.137$. Circles mark the mass bins from 10 to 6, triangles mark the mass bins form 5 to 1. The vertical dashed-line indicates $f_{\mathrm{sub},200} = 0.1$, which is commonly used in the literature as a threshold to divide relaxed and unrelaxed clusters. The Spearman's correlation coefficient is $0.69$.}
\label{f_sub_mass_ratio_figure}
\end{figure}

We also check whether there is a dependence of this ratio on the amount of substructure in the cluster. We report an example plot for quantities evaluated within $R_{200}$ at $z=0.137$ in Figure \ref{f_sub_mass_ratio_figure}. In this particular case, there is a strong correlation between the two quantities, with a Spearman's correlation coefficient of $0.69$. This is true in general for other redshifts and radii, with a value ranging from $0.59$ to $0.76$. There is no particular trend with radius and redshift. What is evident is that, for clusters with low $f_{\mathrm{sub},200}$, there is a large scatter in $f_{\mathrm{sub},200,>1e12}/f_{\mathrm{sub},200}$, while clusters with a high $f_{\mathrm{sub},200}$ also show a higher value of $f_{\mathrm{sub},200,>1e12}/f_{\mathrm{sub},200}$. All the cluster having more than $10 \%$ of their mass in substructures also have more than $70 \%$ of their substructure mass in subhalos more massive than $10^{12} \ {\rm M_{\odot}} \ h^{-1}$. We also detect a population of low mass clusters with a small value of $f_{\mathrm{sub},200}$ but with a large fraction of the mass in substructures residing in massive subhalos.

\begin{figure}
\hbox{
  \epsfig{figure=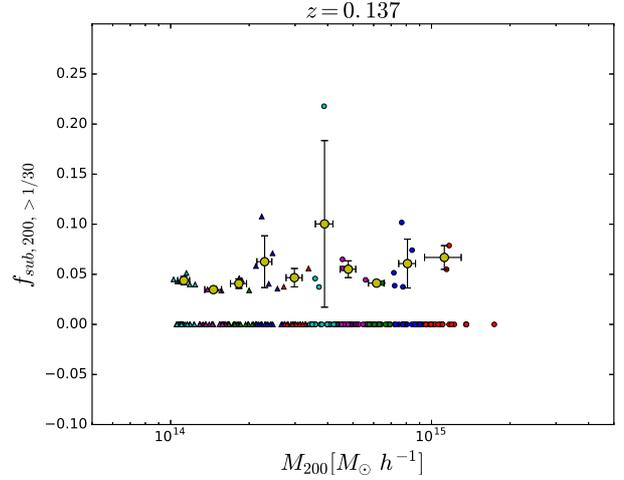,width=0.5\textwidth}
}
\caption{Distribution of $f_{\mathrm{sub},200,>1/30}$ at $z=0.137$. Circles mark the mass bins from 10 to 6, triangles mark the mass bins form 5 to 1. The big yellow circles mark the mean values in every mass bin, considering only the clusters with $f_{\mathrm{sub,vir},>1/30} > 0$. Error bars mark one standard deviation of the mean.}
\label{f_sub_200_30_figure}
\end{figure}

Another analysis is limited to objects with a minimum mass of $M_{\mathrm{vir}}/30$ ($M_{200}/30$, $M_{500}/30$). Those are massive substructures that disturb the virial equilibrium we expect to detect in X-ray images. We thus define the quantity $f_{\mathrm{sub,vir},>1/30}$ ($f_{\mathrm{sub},200,>1/30}$, $f_{\mathrm{sub},500,>1/30}$) as the ratio between the sum of the mass of the satellite halos above the corresponding mass threshold and within $R_{\mathrm{vir}}$ ($R_{200}$, $R_{500}$) and $M_{\mathrm{vir}}$ ($M_{200}$, $M_{500}$). In this case the threshold scales with the enclosed mass, so that self-similarity is preserved. We show an example for $f_{\mathrm{sub},200,>1/30}$ at $z=0.137$ in Figure \ref{f_sub_200_30_figure}. For the great majority of clusters, $f_{\mathrm{sub},200,>1/30}$ is $0$, meaning that the largest part of our sample has no massive SUBFIND substructures within $R_{200}$. Also in this case, there is no correlation with mass. The same consideration is also valid for $R_{\mathrm{vir}}$ and $R_{500}$. We report the results in Table \ref{f_sub_30_table}, where we summarize the fraction of halos with massive substructures and the mean mass fraction in substructures for these halos. The fraction of halos with massive substructures is less than $20 \%$ at any redshift and radius.

\begin{table*}
\begin{center}
\caption{Distribution of $f_{\mathrm{sub,vir},>1/30}$, $f_{\mathrm{sub},200,>1/30}$ and $f_{\mathrm{sub},500,>1/30}$ at different redshifts. We list the percentage of clusters with substructures. The mean values are evaluated only for clusters with substructures.}  \label{f_sub_30_table}
\begin{tabular}{ccccccccccccc} 
\\
\hline
\hline
 & & \multicolumn{3}{c}{$f_{\mathrm{sub,vir},>1/30}$} & & \multicolumn{3}{c}{$f_{\mathrm{sub},200,>1/30}$} && \multicolumn{3}{c}{$f_{\mathrm{sub},500,>1/30}$} \\
\hline
Redshift & & fraction & mean & $\sigma$ & & fraction & mean & $\sigma$ & & fraction & mean & $\sigma$ \\
\hline
$z=0$ & & $14.0\%$ & $0.053$ & $0.019$ & & $18.4 \%$ & $0.048$ & $0.019$ && $16.8 \%$ & $0.054$ & $0.025$ \\
\hline
$z=0.137$ & & $13.6\%$ & $0.065$ & $0.041$ & & $13.2 \%$ & $0.056$ & $0.034$ && $12.8 \%$ & $0.057$ & $0.023$ \\
\hline
$z=0.293$ & & $14.6\%$ & $0.065$ & $0.030$ & & $15.0 \%$ & $0.059$ & $0.031$ && $13.7 \%$ & $0.055$ & $0.031$ \\
\hline
$z=0.470$ & & $15.7\%$ & $0.070$ & $0.045$ & & $13.8 \%$ & $0.065$ & $0.033$ && $12.9 \%$ & $0.053$ & $0.021$ \\
\hline
$z=0.783$ & & $15.7\%$ & $0.066$ & $0.037$ & & $16.3 \%$ & $0.066$ & $0.039$ && $14.5 \%$ & $0.056$ & $0.023$ \\
\hline
$z=1.179$ & & $17.1\%$ & $0.068$ & $0.036$ & & $14.5 \%$ & $0.066$ & $0.033$ && $12.8 \%$ & $0.064$ & $0.038$ \\
\hline
\hline
\end{tabular}
\end{center}
\end{table*}

\noindent We note here that \cite{2007MNRAS.381.1450N} use $f_{\mathrm{sub,vir}} < 0.1$ as one of the three criteria to separate relaxed from unrelaxed clusters, while \cite{2016MNRAS.460.1214L} use $f_{\mathrm{sub},200} < 0.1$. In both cases, their value is about twice our mean and median values. In particular, as it is evident by looking at Figure \ref{f_sub_200_figure}, we only have a very small number of objects with $f_{\mathrm{sub},200} > 0.1$. This indicates that $f_{\mathrm{sub},200} < 0.1$ alone is not a strong criterium to select relaxed clusters.

\subsection{Octants as a new measure of substructures} \label{octants}

In this section we introduce a new method to quantify the mass in substructures within a given radius independent on SUBFIND. The idea is to divide the cluster in octants, and thus we label this methodology as ``octant analysis''. For every cluster, we evaluate the mean radial density profile of total matter in the octants defined by the reference coordinate system of the simulation. The octant density profiles are evaluated in the same linearly-spaced bins of $50 \ {\rm kpc} \ h^{-1}$ we use for the total cluster profile. We also evaluate the $1 \sigma$ standard deviation from the mean radial density profile (of total matter). In every radial bin, in every octant, we define an excess in mass if the density profile is higher than the cluster mean radial density profile plus $1 \sigma$. \footnote{We also check with $2 \sigma$, but we find that this condition is too strong and highly underestimates the number of substructures.} We associate this excess in mass as due to substructures. We evaluate the mass of this substructures in every radial bin of every octant, and by summing them up, we evaluate the total mass in substructures as identified with the octant analysis. For this mass determination, we perform a clipping procedure, {\it i.e.} in every radial bin, we exclude the octants containing an excess when determining the mean value above which we start integrating. We generically indicate the mass fraction in substructures as identified with the octant method with $f_{\mathrm{octants}}$.

We define the substructure fraction within a given overdensity as the ratio between the mass in substructures in the octants within the corresponding radius ($R_{\mathrm{vir}}$, $R_{200}$, $R_{500}$) and the enclosed mass ($M_{\mathrm{vir}}$, $M_{200}$, $M_{500}$). We refer to these three values as $f_{\mathrm{octants,vir}}$, $f_{\mathrm{octants},200}$ and $f_{\mathrm{octants},500}$, respectively. 

\begin{figure*}
\hbox{
  \epsfig{figure=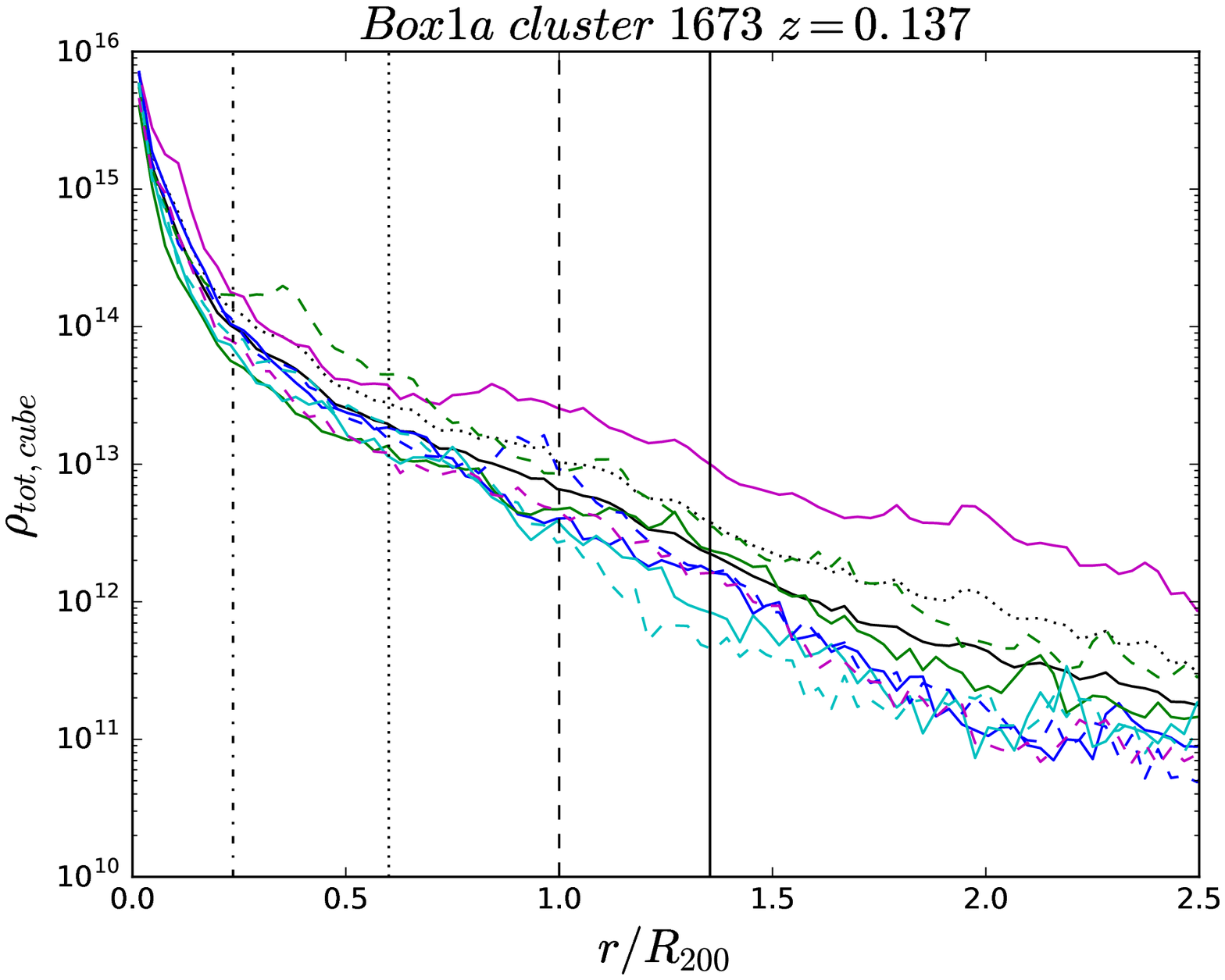,width=0.5\textwidth}
  \epsfig{figure=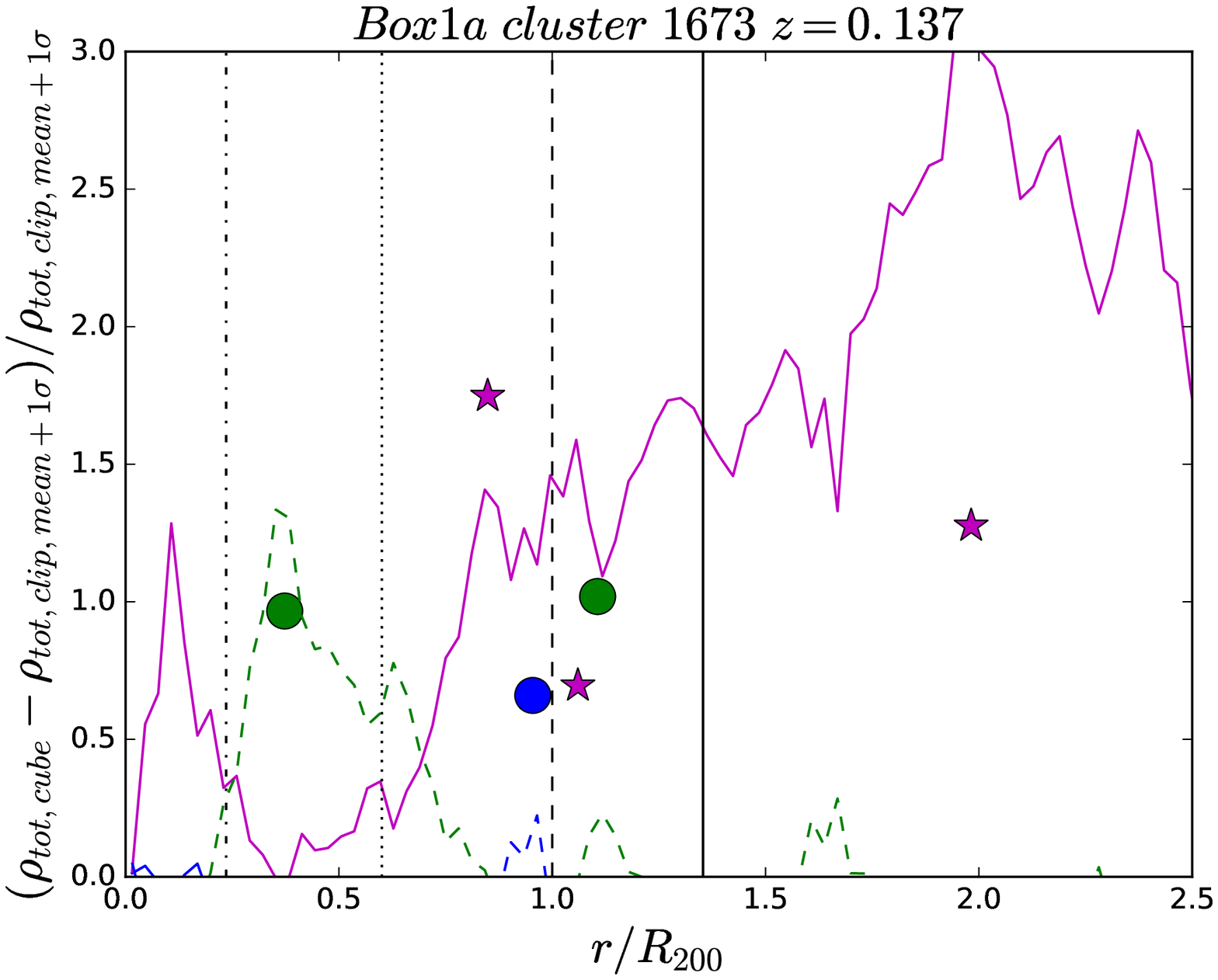,width=0.5\textwidth}
}
\hbox{
  \epsfig{figure=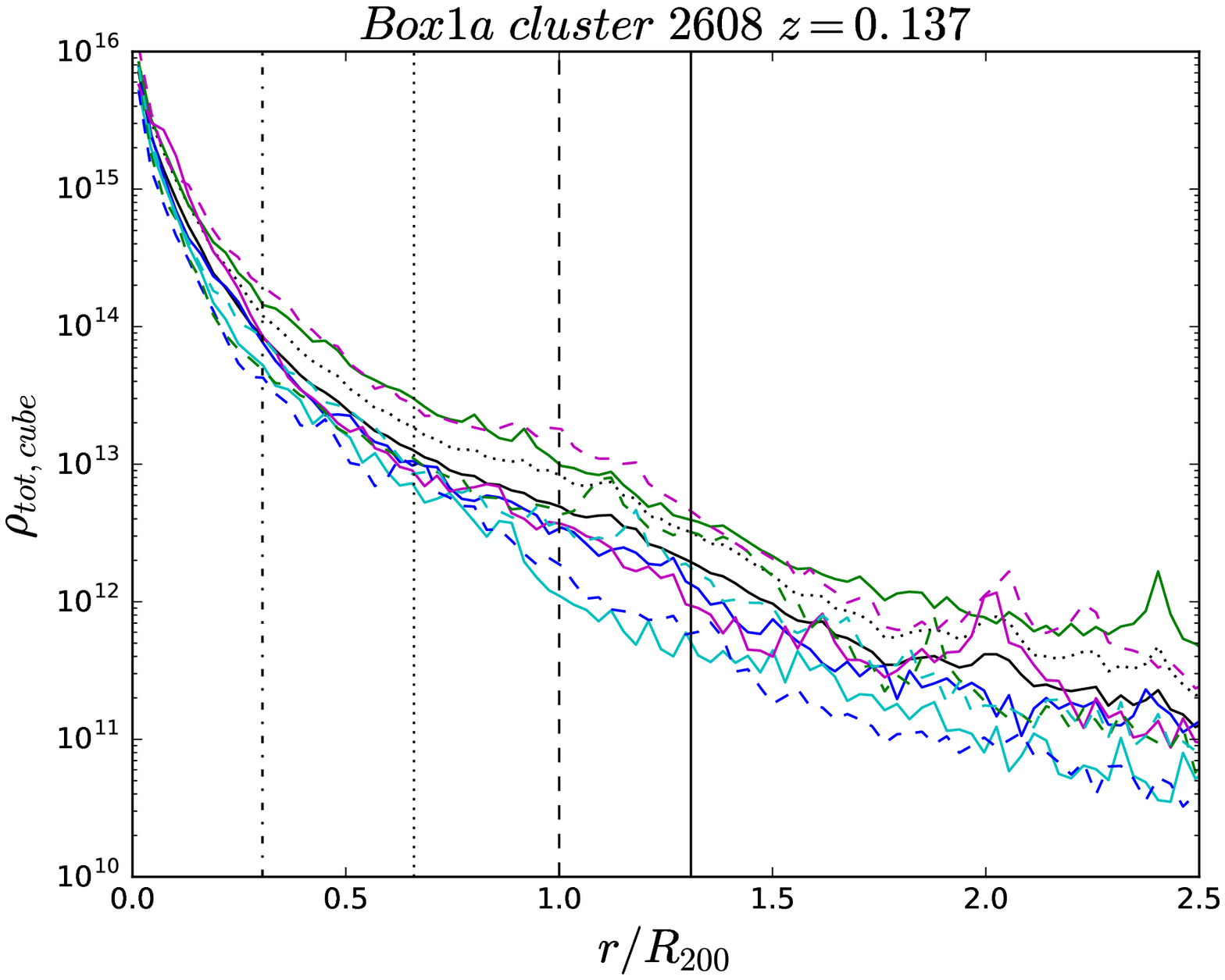,width=0.5\textwidth}
  \epsfig{figure=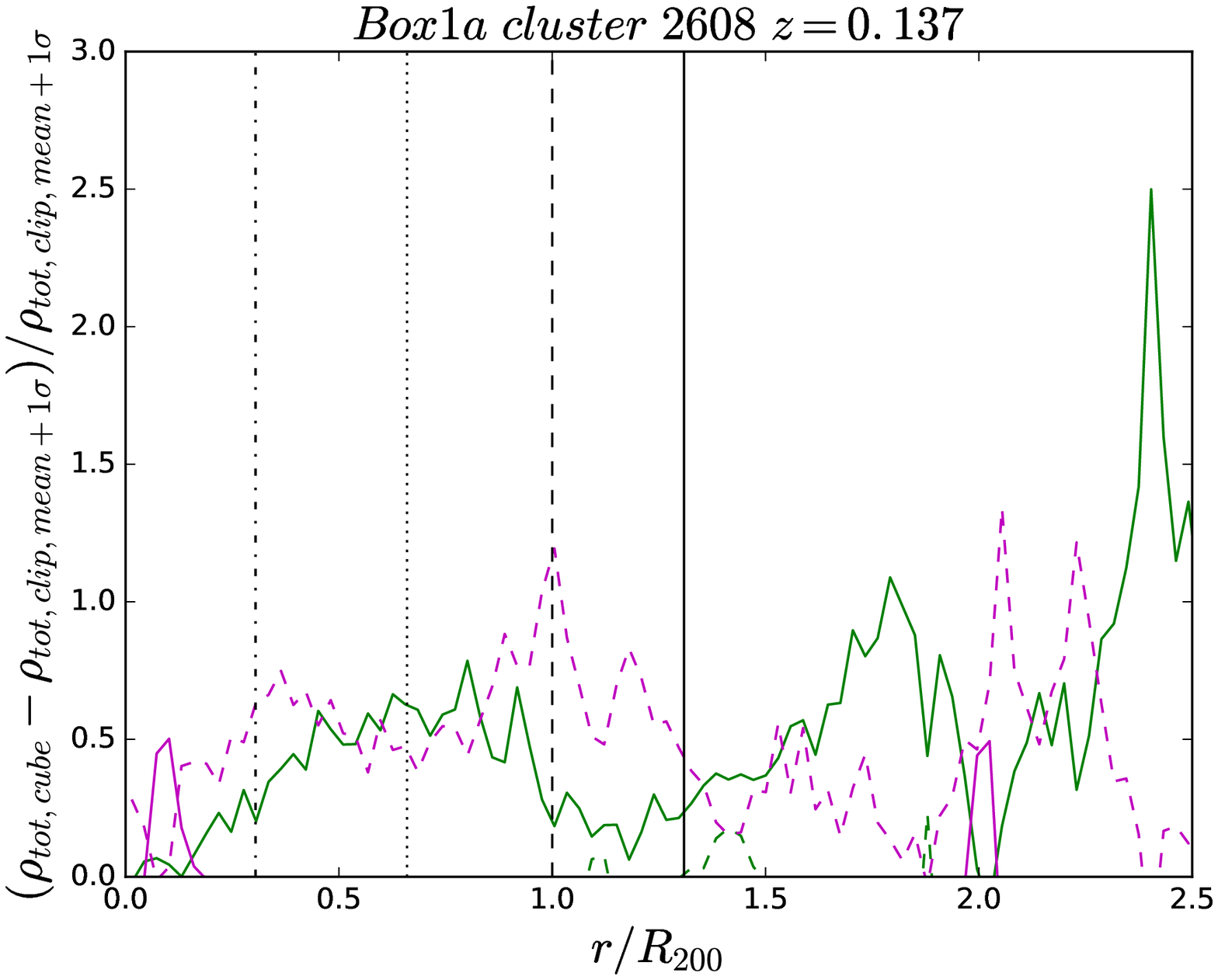,width=0.5\textwidth}
}
\caption{Two examples of the octant method. (Left-panels) Clipped total matter mean radial density (solid black line) and $1 \sigma$ limits (dotted black line). The eight colored lines indicate the total radial density in the octants in which the cluster is divided. The black vertical lines mark $R_{2500}$, $R_{500}$, $R_{200}$ and $R_{\mathrm{vir}}$. (Right-panels) Residuals of the radial density in octants with respect to $1 \sigma$ of the mean radial distribution. The circle and stars mark the location of subhalos more massive than $10^{13} \ {\rm M_{\odot}} \ h^{-1}$ as identified by SUBFIND. The color code is the same as the line, and stars correspond to solid-line octants while circles correspond to dashed-line octants. For the y-axis value of the SUBFIND halos, we use arbitrary units proportional to the mass.}
\label{octants_figure}
\end{figure*}

We show two examples of this procedure in Figure \ref{octants_figure} In the density radial profile of octants we have excess with respect to $1 \sigma$ above the mean, but we never have a deficit with respect to $1 \sigma$ below the mean. This fact tells us that clusters have overdense regions where substructures lie, but they don't have extremely underdense regions and holes. Secondly, we note that to every massive subhalo identified by SUBFIND corresponds an overdense region at the same radial distance from the cluster center, in the same octant. The opposite is not always true. In Figure \ref{octants_figure} we also show an example where there are two overdense regions identified with the octants but there is no substructure identified by SUBFIND.

\begin{figure}
\hbox{
  \epsfig{figure=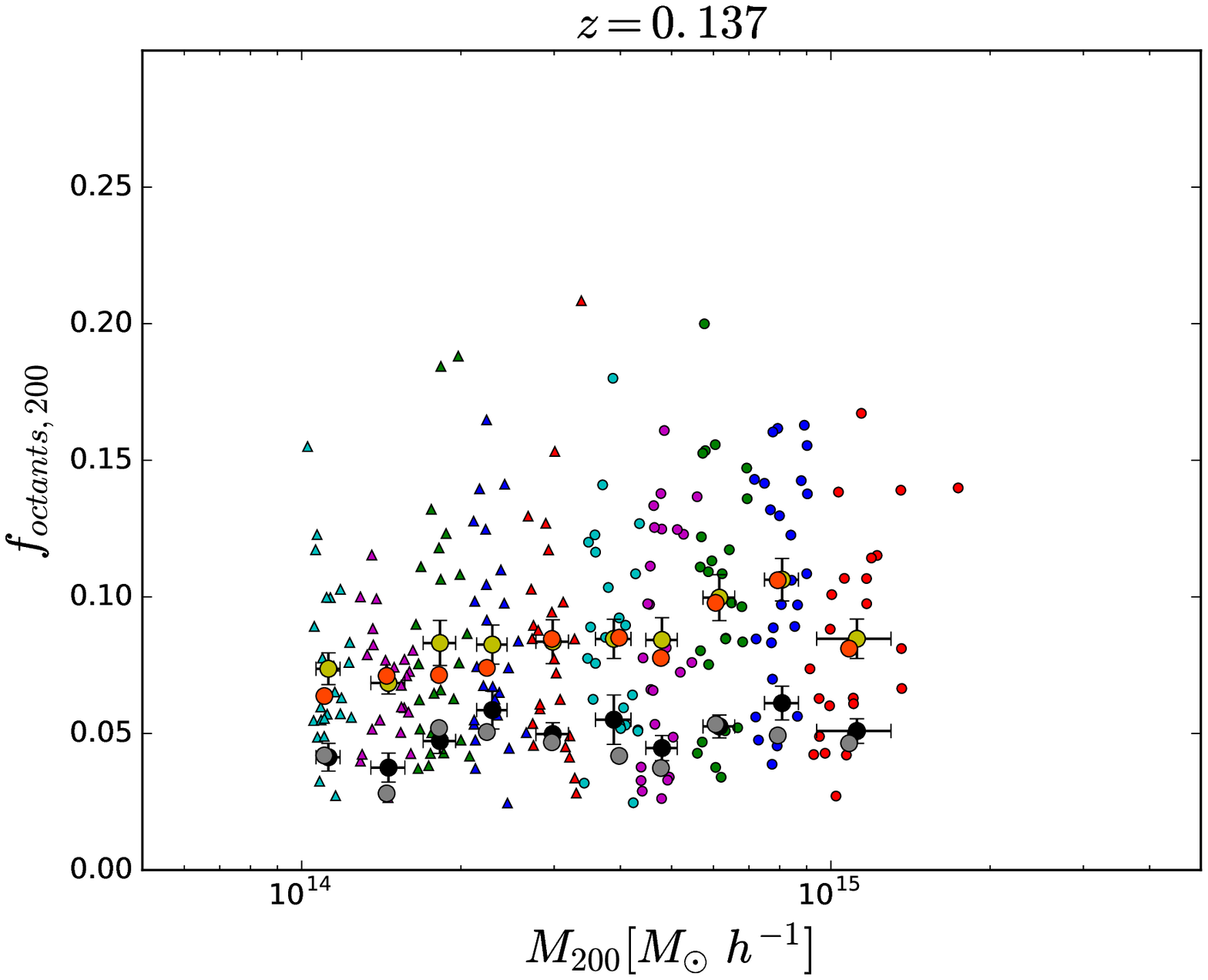,width=0.5\textwidth}
}
\caption{Distribution of $f_{\mathrm{octants},200}$ at $z=0.137$. Circles mark the mass bins from 10 to 6, triangles mark the mass bins form 5 to 1. The big yellow circles mark the mean values in every mass bin, the big orange circles mark the median. Error bars mark one standard deviation of the mean. The big black and grey circles mark the mean and median of $f_{\mathrm{sub},200}$, respectively, as already shown in Figure \ref{f_sub_200_figure}.}
\label{f_octants_200_figure}
\end{figure}

\begin{figure}
\hbox{
  \epsfig{figure=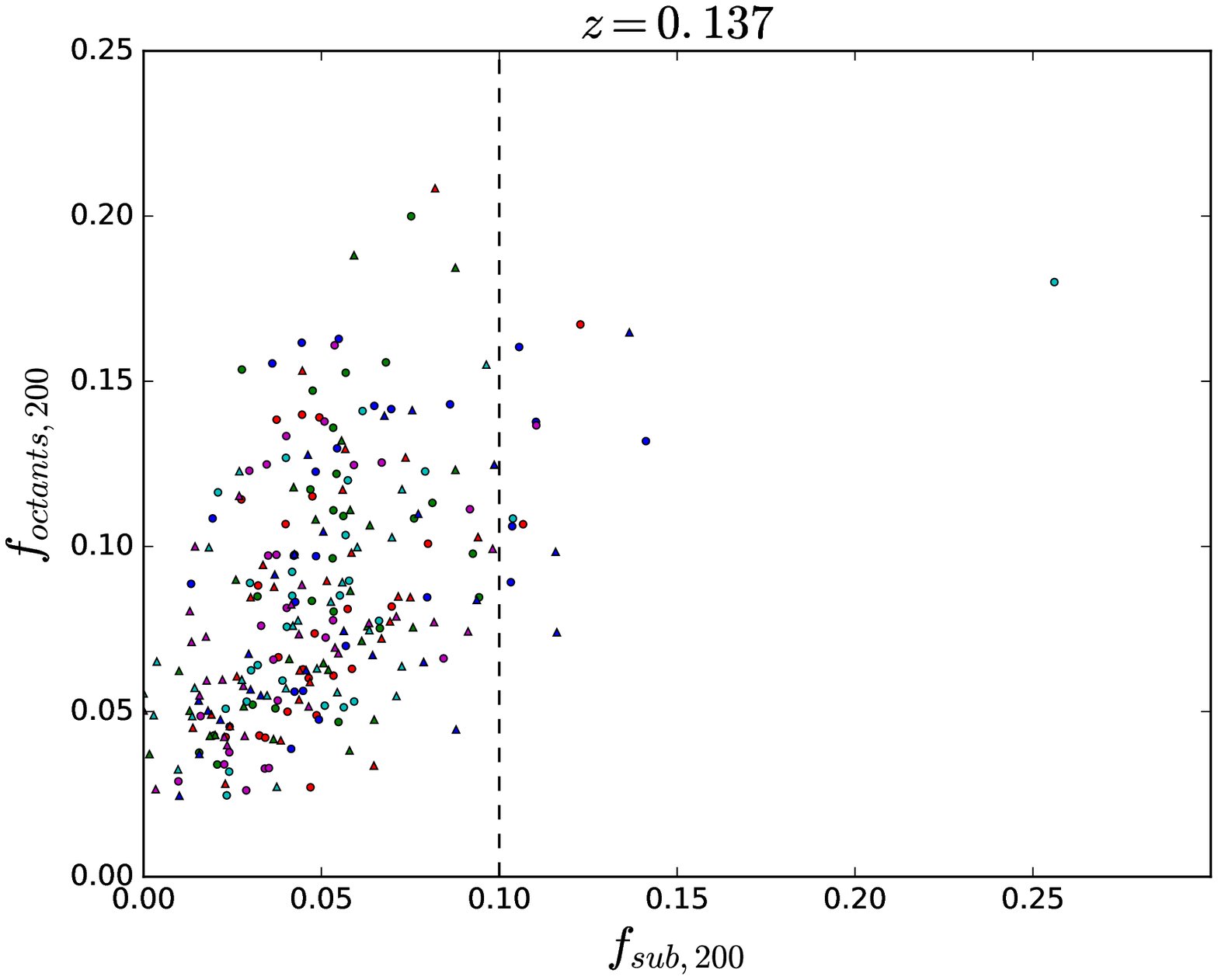,width=0.5\textwidth}
}
\caption{Individual cluster relation between $f_{\mathrm{octants},200}$ and $f_{\mathrm{sub},200}$ at $z=0.137$. Circles mark the mass bins from 10 to 6, triangles mark the mass bins form 5 to 1. The vertical dashed-line indicates $f_{\mathrm{sub},200} = 0.1$, which is commonly used in the literature as a threshold to divide relaxed and unrelaxed clusters. The Spearman's correlation coefficient is $0.53$.}
\label{f_octants-f_sub_200_figure}
\end{figure}

We report an example of the evaluation of the substructure distribution with the octants in Figure \ref{f_octants_200_figure}. It is evident that, as for SUBFIND, also for the fraction in substructures determined with the octants there is only a very weak correlation with mass. We also compare $f_{\mathrm{octants},200}$ with $f_{\mathrm{sub},200}$, that we defined in the previous section. We see a correspondence between massive SUBFIND halos and octant overdensities, but on average the value of  $f_{\mathrm{octants},200}$ is higher than $f_{\mathrm{sub},200}$, because the octant analysis also detects overdensities that have no corresponding SUBFIND halos. On the other hand, there are clusters for which $f_{\mathrm{sub},200}>f_{\mathrm{octants},200}$, because SUBFIND is able to identify small bound substructures not seen in the octant analysis. Indeed, there is only a moderate correlation between the two quantities, as we can see in Figure \ref{f_octants-f_sub_200_figure} where the Spearman's coefficient is $0.53$. The same is true also at other redshifts and at $R_{\mathrm{vir}}$ and $R_{500}$, with a correlation coefficient ranging from $0.39$ to $0.62$ (but this strong correlation shows up only at $z=0$.). We summarize the results for the octants in Table \ref{f_octants_table}.

\begin{table*}
\begin{center}
\caption{Same as Table \ref{f_sub_table}, but for $f_{\mathrm{octants,vir}}$, $f_{\mathrm{octants},200}$ and $f_{\mathrm{octants},500}$.}  \label{f_octants_table}
\begin{tabular}{cccccccccccccccc} 
\\
\hline
\hline
 & \multicolumn{5}{c}{$f_{\mathrm{octants,vir}}$} & \multicolumn{5}{c}{$f_{\mathrm{octants},200}$} & \multicolumn{5}{c}{$f_{\mathrm{octants},500}$} \\
\hline
Redshift & mean & $\sigma$ & median & $16 \%$ & $84 \%$ & \ mean & $\sigma$ & median & $16 \%$ & $84 \%$ & \ mean & $\sigma$ & median & $16 \%$ & $84 \%$ \\
\hline
$z=0$ & $0.085$ & $0.039$ & $0.077$ & $0.045$ & $0.125$ & $\ 0.080$ & $0.039$ & $0.072$ & $0.041$ & $0.118$ & $\ 0.069$ & $0.035$ & $0.061$ & $0.035$ & $0.104$ \\
\hline
$z=0.137$ & $0.092$ & $0.040$ & $0.083$ & $0.054$ & $0.131$ & $\ 0.085$ & $0.038$ & $0.078$ & $0.048$ & $0.125$ & $\ 0.074$ & $0.035$ & $0.066$ & $0.040$ & $0.110$ \\
\hline
$z=0.293$ & $0.094$ & $0.044$ & $0.081$ & $0.051$ & $0.143$ & $\ 0.088$ & $0.043$ & $0.076$ & $0.048$ & $0.133$ & $\ 0.079$ & $0.040$ & $0.069$ & $0.041$ & $0.119$ \\
\hline
$z=0.470$ & $0.096$ & $0.042$ & $0.088$ & $0.056$ & $0.138$ & $\ 0.091$ & $0.038$ & $0.083$ & $0.052$ & $0.136$ & $\ 0.080$ & $0.037$ & $0.071$ & $0.046$ & $0.122$ \\
\hline
$z=0.783$ & $0.102$ & $0.044$ & $0.093$ & $0.058$ & $0.150$ & $\ 0.098$ & $0.042$ & $0.092$ & $0.056$ & $0.142$ & $\ 0.087$ & $0.039$ & $0.078$ & $0.051$ & $0.125$ \\
\hline
$z=1.179$ & $0.110$ & $0.043$ & $0.105$ & $0.069$ & $0.151$ & $\ 0.106$ & $0.042$ & $0.101$ & $0.065$ & $0.141$ & $\ 0.093$ & $0.040$ & $0.087$ & $0.055$ & $0.132$ \\
\hline
\hline
\end{tabular}
\end{center}
\end{table*}

\subsection{Offsets}

In this section we discuss the last method to determine the presence of substructures in a cluster, namely the offset between the potential minimum and the center of mass. To evaluate the center of mass, we consider all the types of particles included in the simulation, and we generally indicate the total offset with $off_{\mathrm{tot}}$. For every cluster, we evaluate the center of mass within $R_{\mathrm{vir}}$ ($R_{200}$, $R_{500}$). We define the offset $off_{\mathrm{tot,vir}}$ ($off_{\mathrm{tot},200}$, $off_{\mathrm{tot},500}$) as the distance between the center of the cluster, identified with the location of the most bound particle, and the center of mass, in units of $R_{\mathrm{vir}}$ ($R_{200}$, $R_{500}$). 

\begin{figure}
\hbox{
  \epsfig{figure=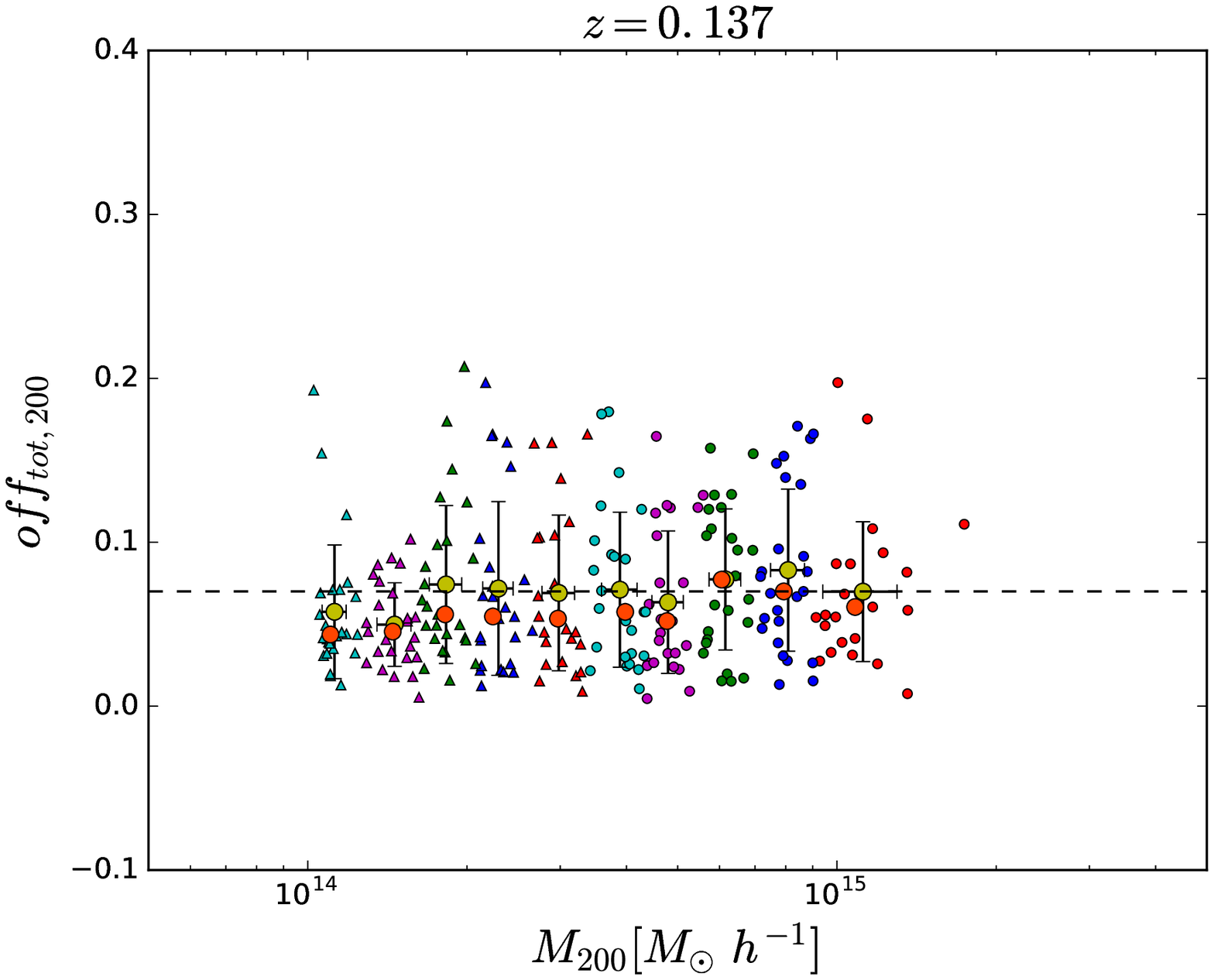,width=0.5\textwidth}
}
\caption{Distribution of $off_{\mathrm{tot},200}$ at $z=0.137$. Circles mark the mass bins from 10 to 6, triangles mark the mass bins form 5 to 1. The big yellow circles mark the mean values in every mass bin, the big orange circles mark the median. Error bars mark one standard deviation of the mean. The horizontal dashed-line indicates $off_{\mathrm{tot},200} = 0.07$, which is commonly used in the literature as a threshold to divide relaxed and unrelaxed clusters.}
\label{off_200_figure}
\end{figure}

We show an example of the distribution of $off_{\mathrm{tot},200}$ for clusters at $z=0.137$ in Figure \ref{off_200_figure}. There is only a very weak correlation with mass at a given redshift. The same considerations are valid also at $R_{\mathrm{vir}}$ and $R_{500}$ so we can proceed by evaluating the average values within a given snapshot independently on mass. Results for $off_{\mathrm{tot,vir}}$, $off_{\mathrm{tot},200}$ and $off_{\mathrm{tot},500}$ are reported in Table \ref{off_table}.

\begin{table*}
\begin{center}
\caption{Same as Table \ref{f_sub_table}, but for $off_{\mathrm{tot,vir}}$, $off_{\mathrm{tot},200}$ and $off_{\mathrm{tot},500}$.}  \label{off_table}
\begin{tabular}{cccccccccccccccc} 
\\
\hline
\hline
 & \multicolumn{5}{c}{$off_{\mathrm{tot,vir}}$} & \multicolumn{5}{c}{$off_{\mathrm{tot},200}$} & \multicolumn{5}{c}{$off_{\mathrm{tot},500}$} \\
\hline
Redshift & mean & $\sigma$ & median & $16 \%$ & $84 \%$ & \ mean & $\sigma$ & median & $16 \%$ & $84 \%$ & \ mean & $\sigma$ & median & $16 \%$ & $84 \%$ \\
\hline
$z=0$ & $0.060$ & $0.041$ & $0.051$ & $0.023$ & $0.097$ & $\ 0.065$ & $0.048$ & $0.052$ & $0.022$ & $0.113$ & $\ 0.063$ & $0.048$ & $0.049$ & $0.021$ & $0.102$ \\
\hline
$z=0.137$ & $0.067$ & $0.045$ & $0.054$ & $0.027$ & $0.108$ & $\ 0.069$ & $0.046$ & $0.055$ & $0.026$ & $0.120$ & $\ 0.065$ & $0.044$ & $0.055$ & $0.023$ & $0.105$ \\
\hline
$z=0.293$ & $0.072$ & $0.046$ & $0.059$ & $0.030$ & $0.119$ & $\ 0.074$ & $0.051$ & $0.059$ & $0.029$ & $0.123$ & $\ 0.074$ & $0.055$ & $0.055$ & $0.031$ & $0.129$ \\
\hline
$z=0.470$ & $0.075$ & $0.049$ & $0.062$ & $0.032$ & $0.123$ & $\ 0.076$ & $0.049$ & $0.062$ & $0.033$ & $0.126$ & $\ 0.077$ & $0.051$ & $0.063$ & $0.031$ & $0.134$ \\
\hline
$z=0.783$ & $0.077$ & $0.046$ & $0.072$ & $0.031$ & $0.122$ & $\ 0.080$ & $0.048$ & $0.069$ & $0.034$ & $0.129$ & $\ 0.082$ & $0.053$ & $0.072$ & $0.035$ & $0.138$ \\
\hline
$z=1.179$ & $0.083$ & $0.048$ & $0.077$ & $0.035$ & $0.136$ & $\ 0.084$ & $0.048$ & $0.075$ & $0.035$ & $0.137$ & $\ 0.088$ & $0.051$ & $0.075$ & $0.039$ & $0.144$ \\
\hline
\hline
\end{tabular}
\end{center}
\end{table*}

\noindent With respect to the literature, we see that the value $off_{\mathrm{tot},200}=0.07$ (or equivalently $off_{\mathrm{tot,vir}}=0.07$) is a reasonable threshold to divide relaxed and unrelaxed clusters, but it is not equally good at all redshifts.

\subsection{Comparison of methods} \label{method_comparison}

\noindent We discuss the strengths and weaknesses of the methods we introduced before in this section, and in particular their usefulness in an observational perspective.

SUBFIND is able to identify smaller bound substructures better than the octants, but does not detect diffuse overdensities. Its dynamical range allows for lower values with respect to the octant method. Moreover, if we concentrate only on massive satellite halos, we see that the great majority of the halos is not marked to have a substructure. This is a critical result from the observational perspective, because massive substructures in the central regions are the ones that we expect to observe in X-rays images, and SUBFIND puts a very stringent constraint on their number.

The octant method that we introduced in Section \ref{octants} is able to fully identify all the bound substructures in very substructured clusters, but also account for other excess in mass that is not found by SUBFIND. Its dynamical range allows for larger values with respect to SUBFIND. Indeed, the octant method returns on average a higher subhalo mass fraction with respect to SUBFIND. There is a moderate correlation between the two quantities.

Finally, the offset is in principle the most straightforward method, because it is based only on the particle distribution, without other binding or density threshold condition. In principle, however, the offset is more a measure of the geometrical distribution of matter than an estimate of the presence of substructures.

\begin{figure*}
\hbox{
  \epsfig{figure=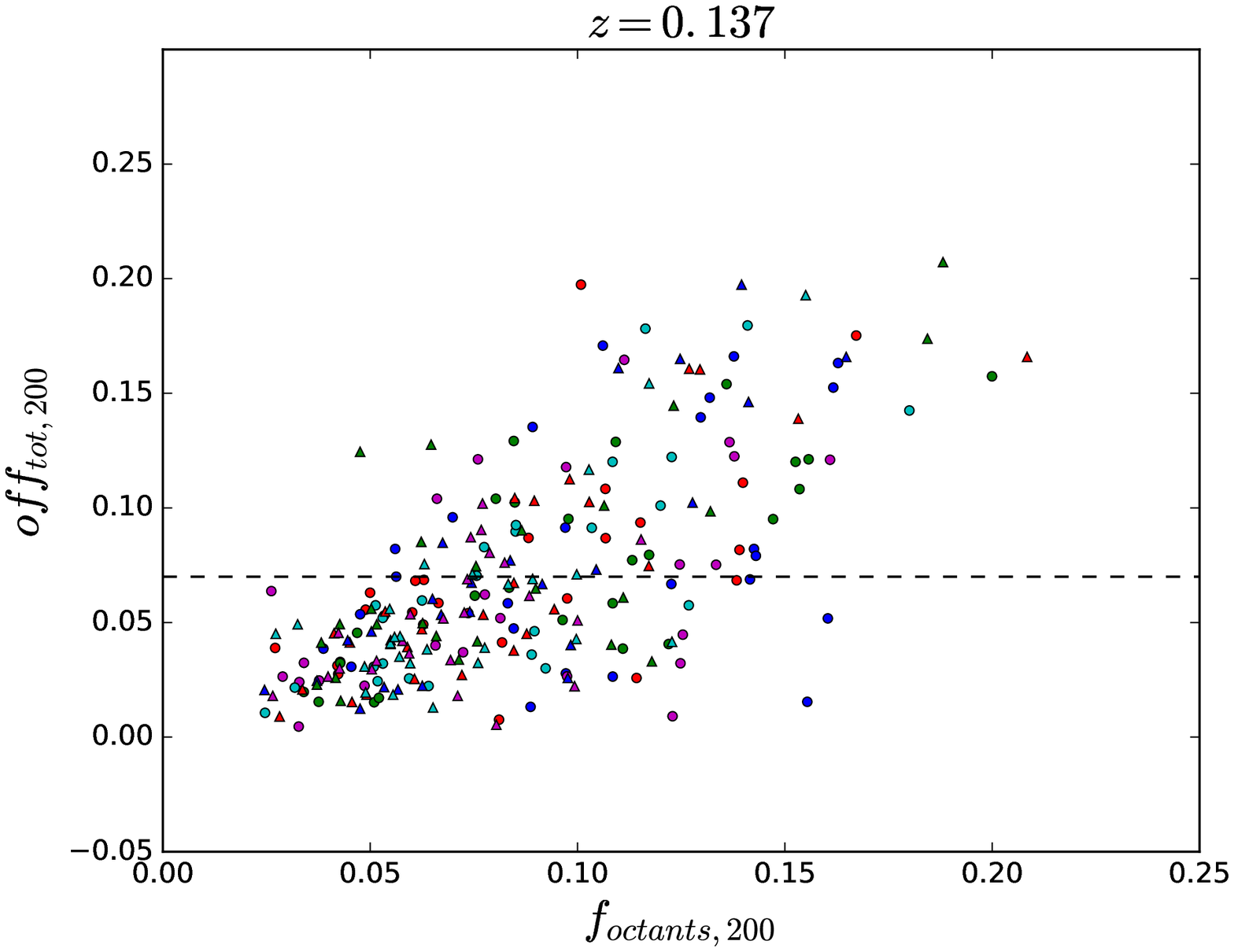,width=0.5\textwidth}
  \epsfig{figure=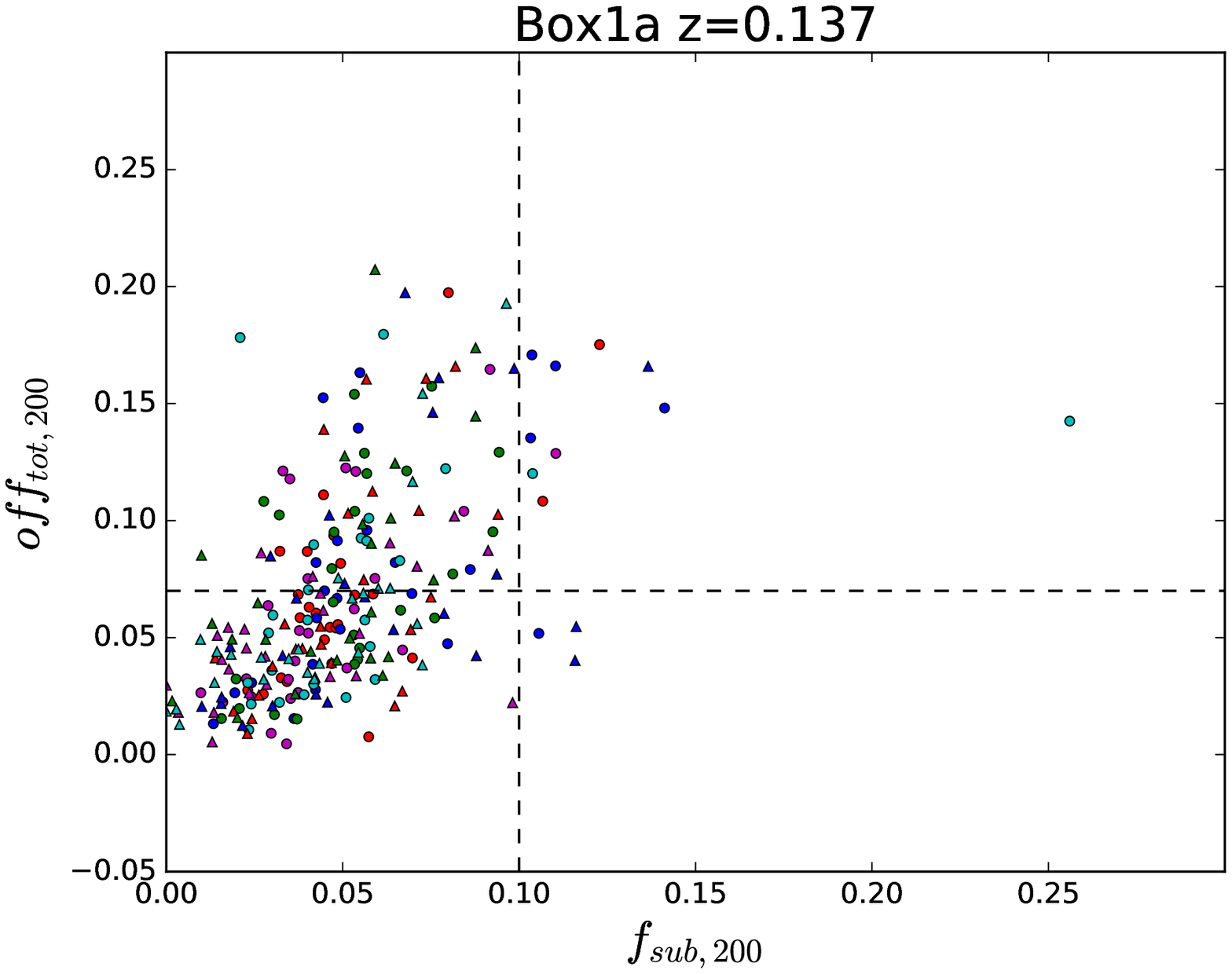,width=0.5\textwidth}
}
\caption{(Left-panel) Individual cluster relation between $off_{\mathrm{tot},200}$ and $f_{\mathrm{octants},200}$ at $z=0.137$. Circles mark the mass bins from 10 to 6, triangles mark the mass bins form 5 to 1. The horizontal dashed-line indicates $off_{\mathrm{tot,vir}} = 0.07$, which is commonly used in the literature as a threshold to divide relaxed and unrelaxed clusters. The Spearman's correlation coefficient is $0.65$. (Right-panel) Same as left-panel, but for the individual cluster relation between $off_{\mathrm{tot},200}$ and $f_{\mathrm{sub},200}$ at $z=0.137$. The vertical dashed-line indicates $f_{\mathrm{sub},200} = 0.1$, which is commonly used in the literature as a threshold to divide relaxed and unrelaxed clusters. The Spearman's correlation coefficient is $0.62$. Adding the $f_{\mathrm{sub},200}<0.1$ condition to the $off_{\mathrm{tot},200}<0.07$ to separate relaxed and unrelaxed cluster only excludes $3$ clusters from the relaxed sample.}
\label{off-octants_scatter}
\end{figure*}

We already pointed out that, in principle, the offset is just a measure of the geometrical distribution of matter within a given radius, but the question whether it is also a measure of the quantity of substructure naturally arises. In particular, it is interesting to investigate the individual cluster relation between the substructure fraction identified by the octants and the value of the offset. As a reference case, we consider $f_{\mathrm{octants},200}$ and $off_{\mathrm{tot},200}$ at $z=0.137$, and we show their relation in Figure \ref{off-octants_scatter}. The statistical test indicates a strong correlation between the two quantities, with a Spearman's coefficient of $0.65$. This result is true also at other redshifts and for other radii, with a Spearman's coefficient ranging from $0.60$ to $0.74$. This is interesting because on the one hand there is no reason that guarantees that to a high offset also has to correspond a high mass fraction in substructures (and vice versa). On the other hand, however, we already saw that the frequency of halos with a big substructure is low, so all the halos with a high $f_{\mathrm{octants},200}$ also have a high $off_{\mathrm{tot},200}$ because there is only one big substructure contributing to the mass fraction, and this substructure is also responsible of the offset.

\noindent In the right panel of Figure \ref{off-octants_scatter}, we show an example of the correlation between $f_{\mathrm{sub},200}$ and $off_{\mathrm{tot},200}$ at $z=0.137$. Depending on redshift and radius, the correlation ranges from moderate to strong, with a Spearman's coefficient ranging from $0.54$ to $0.72$, but it is always less pronounced than the one between $f_{\mathrm{octants}}$ and $off_{\mathrm{tot}}$. This was also the case for the correlation between $f_{\mathrm{sub}}$ and $f_{\mathrm{octants}}$ (see Figure \ref{f_sub_mass_ratio_figure}). As we mentioned before, there is a substantial difference in the behavior of $f_{\mathrm{sub}}$ with respect to $f_{\mathrm{octants}}$ and $off_{\mathrm{tot}}$.

\section{Mass, redshift and radial substructure evolution} \label{evolution}

None of the methods presented in the previous section predicts any strong correlation with mass at fixed redshift. So, at any redshift, the maximum evolution we can expect with mass is given by the intrinsic scatter of the distribution.

\noindent For this reason, it is interesting to study the redshift evolution of substructures. We provide several fits of the different quantities, namely: 1) a constant value; 2) a linear function, $y = y_{0,l} + s(1+z)$; 3) a power law, $y = y_{0,p}(1+z)^a$. We calculate errors by $\sigma / \sqrt{N-1}$, with $N$ being the number of clusters, and the best-fit values are shown in Table \ref{redshift_evolution}, along with $\tilde{\chi}^2$.

\begin{table*}
\begin{center}
\caption{Best-fit parameters for constant, linear function and power law fit to the evolution of various quantities with redshift.} \label{redshift_evolution}
\begin{tabular}{ccccccccccccccccc}
\hline
\hline
 & & \multicolumn{3}{c}{Constant} & & \multicolumn{5}{c}{Linear} & & \multicolumn{5}{c}{Power law} \\
& & $y_{0}$ & $\sigma_{y_{0}}$ & $\tilde{\chi}^2$ & & $y_{0,l}$ & $\sigma_{y_{0,l}}$ & $s$ & $\sigma_s$ & $\tilde{\chi}^2$ & & $y_{0,p}$ & $\sigma_{y_{0,p}}$ & $a$ & $\sigma_a$ & $\tilde{\chi}^2$ \\
\hline
$f_{\mathrm{sub,vir}}$ & & $0.058$ & $0.001$ & $0.41$ & & $0.058$ & $0.004$ & $-0.000$ & $0.002$ & $0.50$ & & $0.058$ & $0.001$ & $-0.003$ & $0.062$ & $0.51$ \\
\hline
$f_{\mathrm{sub,vir},>1e12}$ & & $0.045$ & $0.001$ & $0.28$ & & $0.045$ & $0.004$ & $+0.000$ & $0.002$ & $0.35$ & & $0.045$ & $0.001$ & $+0.016$ & $0.085$ & $0.35$ \\
\hline
$f_{\mathrm{sub},200}$ & & $0.052$ & $0.001$ & $1.18$ & & $0.044$ & $0.004$ & $+0.006$ & $0.003$ & $0.53$ & & $0.049$ & $0.001$ & $+0.172$ & $0.083$ & $0.42$ \\
\hline
$f_{\mathrm{sub},200,>1e12}$ & & $0.041$ & $0.001$ & $0.81$ & & $0.036$ & $0.004$ & $+0.004$ & $0.003$ & $0.52$ & & $0.039$ & $0.001$ & $+0.139$ & $0.093$ & $0.45$ \\
\hline
$f_{\mathrm{sub},500}$ & & $0.039$ & $0.001$ & $2.35$ & & $0.026$ & $0.004$ & $+0.010$ & $0.003$ & $0.21$ & & $0.036$ & $0.001$ & $+0.360$ & $0.105$ & $0.15$ \\
\hline
$f_{\mathrm{sub},500,>1e12}$ & & $0.031$ & $0.001$ & $2.49$ & & $0.020$ & $0.004$ & $+0.008$ & $0.002$ & $0.26$ & & $0.028$ & $0.001$ & $+0.381$ & $0.110$ & $0.19$ \\
\hline
$f_{\mathrm{octants,vir}}$ & & $0.094$ & $0.001$ & $8.82$ & & $0.067$ & $0.004$ & $+0.020$ & $0.003$ & $0.54$ & & $0.086$ & $0.002$ & $+0.307$ & $0.046$ & $0.40$ \\
\hline
$f_{\mathrm{octants},200}$ & & $0.088$ & $0.001$ & $10.11$ & & $0.060$ & $0.004$ & $+0.022$ & $0.003$ & $0.22$ & & $0.081$ & $0.001$ & $+0.344$ & $0.047$ & $0.14$ \\
\hline
$f_{\mathrm{octants},500}$ & & $0.077$ & $0.001$ & $9.36$ & & $0.050$ & $0.004$ & $+0.020$ & $0.003$ & $0.58$ & & $0.040$ & $0.001$ & $+0.371$ & $0.053$ & $0.38$ \\
\hline
$off_{\mathrm{tot,vir}}$ & & $0.071$ & $0.001$ & $5.67$ & & $0.046$ & $0.005$ & $+0.018$ & $0.004$ & $1.10$ & & $0.064$ & $0.002$ & $+0.367$ & $0.072$ & $0.81$ \\
\hline
$off_{\mathrm{tot},200}$ & & $0.073$ & $0.001$ & $4.11$ & & $0.052$ & $0.001$ & $+0.016$ & $0.004$ & $0.44$ & & $0.067$ & 0.002$$ & $+0.314$ & $0.070$ & $0.32$ \\
\hline
$off_{\mathrm{tot},500}$ & & $0.072$ & $0.001$ & $6.78$ & & $0.041$ & $0.006$ & $+0.023$ & $0.004$ & $0.70$ & & $0.064$ & $0.002$ & $+0.446$ & $0.077$ & $0.52$ \\
\hline
\hline
\end{tabular}
\end{center}
\end{table*}

\noindent We start our analysis with the SUBFIND method. The redshift evolution of $f_{\mathrm{sub,vir}}$ is compatible with a constant value of about $6 \%$. A linear relation or a power law do not provide a better fit. Similar considerations can be done for $f_{\mathrm{sub,vir},>1e12}$, with the only difference that the constant value in this case is $4.5 \%$. This means that the contribution of small satellites is constant with redshift, which was already indicated in Table \ref{f_sub_mass_ratio_table}. Contrary to $f_{\mathrm{sub,vir}}$, $f_{\mathrm{sub},200}$ and $f_{\mathrm{sub},500}$ are better described by a linear relation or a power law than by a constant value, even if the evolution is mild: the best-fit linear slope is $0.006 \pm 0.003$ for $f_{\mathrm{sub},200}$ and  $0.010 \pm 0.003$ for $f_{\mathrm{sub},500}$. The normalization is lower than for smaller radii. Correspondingly, also $f_{\mathrm{sub},200,>1e12}$ and $f_{\mathrm{sub},400,>1e12}$ are described by a fit that allows an evolution with redshift.

\noindent In the case of the octant analysis, there is a strong positive correlation between $f_{\mathrm{octants,vir}}$ and redshift, not compatible with a constant value. Both the linear and the power law fitting formulae provide similar $\tilde{\chi}^2$ with a positive trend with redshift. For the linear relation, the slope is $0.020 \pm 0.003$. This result is in tension to what we found for $f_{\mathrm{sub,vir}}$: the SUBFIND method returns a constant fraction of mass in substructures through all the redshift range, while the octants detect an increasing mass fraction. $f_{\mathrm{octants},200}$ and $f_{\mathrm{octants},500}$ have a similar redshift evolution, with a linear slope of $0.022 \pm 0.003$ and $0.020 \pm 0.003$, respectively. Also in this case, the normalization is lower for smaller radii.

\noindent Finally, we consider the redshift evolution of the offset. $off_{\mathrm{tot,vir}}$ has a strong positive correlation with redshift, not compatible with a constant value. The linear slope is $0.018 \pm 0.04$. The evolution of $off_{\mathrm{tot},200}$ and $off_{\mathrm{tot},500}$ is compatible with the one of $off_{\mathrm{tot,vir}}$: the linear slopes are $0.016 \pm 0.004$ and $0.023 \pm 0.004$, respectively, and the the normalization is similar.
 
\noindent We noted earlier in \ref{methods} that the mean values were driven by a set of objects with higher mass fraction or offset, resulting in values significantly higher than the median. We assess this problem here, limiting our study to the octant and offset methods within the virial radius, which are the more interesting from an observational perspective. This time we analyze $f_{\mathrm{octants,vir}}$ and $off_{\mathrm{tot,vir}}$ without considering the objects with values above the mean value plus one standard deviation. In this way we exclude the extreme cases that bias the results. We find that the strong positive correlation with redshift is preserved for both quantities. Also, the slope is identical while of course the normalization is lower, and the effect of removing the outliers on the normalization is similar for $f_{\mathrm{octants,vir}}$ and for $f_{\mathrm{sub,vir}}$.

\noindent Comparing the evolution characterized by the three methods, we see that while $f_{\mathrm{sub,vir}}$ is constant with redshift, both $f_{\mathrm{sub},200}$ and $f_{\mathrm{sub},500}$ have a positive trend with redshift, even if not extremely pronounced ($2 \sigma$ and $3 \sigma$ from a constant value, repsectively). That means that at $z=0$ the inner regions of clusters are less substructured with respect to the virial region, while at $z=1.179$ the values are similar.

This is not the case of $f_{\mathrm{octants}}$, for which the redshift evolution is the same at the three radii considered and is more than $5 \sigma$ from a constant value. According to this method, the inner regions, in particular the one within $R_{500}$, have a lower mass fraction in substructures than the virial region, at all redshifts.

For the offset, not only the redshift evolution is similar at the three radii (more than $4 \sigma$ from a constant value), but also the normalization is the same. 

\begin{figure}
\hbox{
  \epsfig{figure=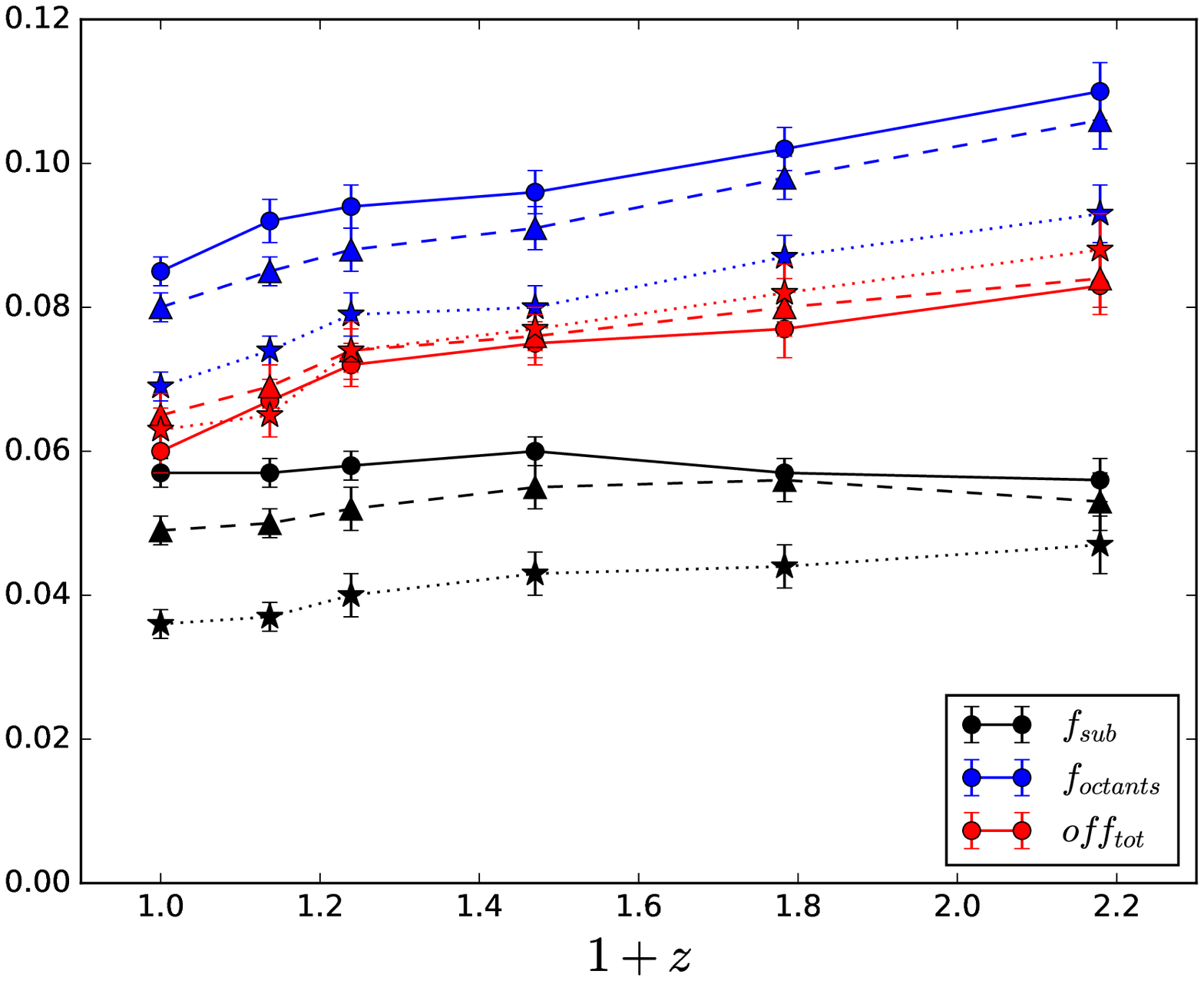,width=0.5\textwidth}
}
\caption{Evolution with redshift of the mean values of $f_{\mathrm{sub}}$ (black), $f_{\mathrm{octants}}$ (blue) and $off_{\mathrm{tot}}$ (red) at different radii: $R_{\mathrm{vir}}$ (solid line and dots), $R_{200}$ (dashed line and triangles) and $R_{500}$ (dotted line and stars).}
\label{sub_evo_compilation}
\end{figure}

We sum up all the previous considerations on the redshift and radial evolution of the diagnostics that we use to determine the distribution of substructures in Figure \ref{sub_evo_compilation}.

\section{Discussion and conclusions} \label{conclusions}

In this paper we studied the evolution of cluster substructure in the mass distribution as a function of redshift, mass and radius. We use three different approaches to address the substructure properties: 1) the SUBFIND halos; 2) the octant method; 3) the offset between the potential minimum and the center of mass. The octant method is a new technique that we introduce here for the first time.

We used a sample of $1226$ simulated clusters, at six different redshifts spanning the range $z=0 - 1.179$ and in ten mass bins in the range $M_{200} = 10^{14} - 1.74 \times 10^{15} \ {\rm M_{\odot}} \ h^{-1}$, and studied in detail the distribution of substructures. 

\noindent Here we recap the main findings of this work:

\begin{itemize}

\item We find that the mass fraction in substructures as seen by the standard SUBFIND method, $f_{\mathrm{sub}}$, does not correlate strongly with mass, at every radius and redshift. $f_{\mathrm{sub,vir}}$ does not show significant evolution with redshift and its value is close to $6 \%$, while $f_{\mathrm{sub},200}$ and $f_{\mathrm{sub},500}$ only show a moderate positive redshift evolution, with a general lower value than $f_{\mathrm{sub,vir}}$.

\item By limiting the analysis to satellite halos more massive than $10^{12} \ {\rm M_{\odot}} \ h^{-1}$, we find that their mean contribution on the subhalo mass fraction of individual clusters is at most $30 \%$, independently of redshift and with no strong correlation with mass, at $R_{\mathrm{vir}}$ and $R_{200}$. At $R_{500}$ their contribution can reach almost $40 \%$, which is not surprising, since in this case we are normalizing by $M_{500}$ which is more sensitive to $10^{12} \ {\rm M_{\odot}} \ h^{-1}$ halos.

\item Analyzing only massive subhalos within a given radius, namely objects more massive than $1/30$ of the enclosed mass, we find that, at all redshifts and radii, less than $20 \%$ of the halos have at least one of these massive substructures. The mean value of the mass in substructures for this subset is higher than the corresponding mean value of $f_{\mathrm{sub}}$, in particular at high redshift and small radii. This result puts tight constraints on the fraction of observed clusters with massive substructures.

\item Unlike SUBFIND, our new octant method finds a strong positive correlation between a mass excess relative to the mean profile and redshift, statistically incompatible with no redshift evolution also at $R_{\mathrm{vir}}$. Notably, the best-fit slope of the redshift evolution is comparable at the three radii considered. The normalization increases with increasing radius. $f_{\mathrm{octants}}$ is on average higher than $f_{\mathrm{sub}}$ and the difference increases at higher redshift. At all redshifts and radii, there is a moderate correlation between $f_{\mathrm{octants}}$ and $f_{\mathrm{sub}}$ for individual clusters.

\item Also the offset between the potential minimum and the center of mass has a strong positive correlation with redshift incompatible with no redshift evolution. For this quantity, both the best-fit slope and normalization are compatible at the different radii considered. The mean and median values are lower than $off_{\mathrm{tot}} = 0.07$ (the value usually quoted as the threshold between relaxed and unrelaxed halos) at low redshift while they are higher at high redshift, indicating an effective evolution of the cluster morphology.

\item We find that there is a strong statistical correlation between individual cluster $f_{\mathrm{octants}}$ and $off_{\mathrm{tot}}$, at all redshift and radii. This is not obvious {\it a priori}, and it is an interesting result. We attribute this strong correlation to the halos that have a big substructure within the enclosed radius. The presence of a massive substructure results in both a high mass fraction and a high offset, thus leading to a correlation between the two quantities.

\end{itemize}

\noindent The first result is that none of the different methods used to quantify the presence of substructures correlates strongly with mass. Mass is the key physical quantity to characterize galaxy clusters as cosmological tools, and we find that the subhalo distribution and evolution are independent of it.

The SUBFIND subhalo mass fraction shows no significant evolution with redshift at $R_{\mathrm{vir}}$ while it increases with redshift at $R_{200}$ and $R_{500}$. The maximum evolution between $z=0$ and $z=1.179$ is $5 \%$, $14 \%$ and $23 \%$, respectively.
Our new method based on the octants finds a mass excess with respect to the mean profile that increases with redshift. The increase from $z=0$ to $z=1.179$ ranges from $23 \%$ at $R_{\mathrm{vir}}$ to $26 \%$ at $R_{500}$.
Interestingly, the redshift evolution of $f_{\mathrm{octants}}$ is similar to the one of $off_{\mathrm{tot}}$, showing an increase from $z=0$ to $z=1.179$ of $28 \%$ at $R_{\mathrm{vir}}$ and $R_{500}$ and of $23 \%$ at $R_{200}$. We also find a strong correlation between these two quantities, on an individual cluster basis. This result deserves further investigation, but suggests these two indicators are very useful to characterize the presence of substructures in a cluster, also from an observational perspective.

\section*{Acknowledgements}

We acknowledge support from the DFG through the Munich Excellence Cluster ``Structure and Evolution of the Universe'', and H.B. and G.C. thank for the support from the DFG Transregio Program TR33.
K.D. acknowledges the support by the DFG Cluster of Excellence ``Origin and Structure
of the Universe''. We are especially grateful for the support by M. Petkova
through the Computational Center for Particle and Astrophysics (C2PAP). Computations
have been performed at the ``Leibniz-Rechenzentrum'' with CPU time assigned to the
Project ``pr86re''.
Information on the Magneticum Pathfinder project is available at \url{http://www.magneticum.org}.
This research made use of Astropy, a community-developed core Python package for Astronomy \citep{2013A&A...558A..33A}, Matplotlib library \citep{Hunter:2007} and pyGadgetReader code \citep{2014ascl.soft11001T}.





\bibliographystyle{mnras} 
\bibliography{cristiano} 

\begin{thebibliography}{}
\makeatletter
\relax
\def\mn@urlcharsother{\let\do\@makeother \do\$\do\&\do\#\do\^\do\_\do\%\do\~}
\def\mn@doi{\begingroup\mn@urlcharsother \@ifnextchar [ {\mn@doi@}
  {\mn@doi@[]}}
\def\mn@doi@[#1]#2{\def\@tempa{#1}\ifx\@tempa\@empty \href
  {http://dx.doi.org/#2} {doi:#2}\else \href {http://dx.doi.org/#2} {#1}\fi
  \endgroup}
\def\mn@eprint#1#2{\mn@eprint@#1:#2::\@nil}
\def\mn@eprint@arXiv#1{\href {http://arxiv.org/abs/#1} {{\tt arXiv:#1}}}
\def\mn@eprint@dblp#1{\href {http://dblp.uni-trier.de/rec/bibtex/#1.xml}
  {dblp:#1}}
\def\mn@eprint@#1:#2:#3:#4\@nil{\def\@tempa {#1}\def\@tempb {#2}\def\@tempc
  {#3}\ifx \@tempc \@empty \let \@tempc \@tempb \let \@tempb \@tempa \fi \ifx
  \@tempb \@empty \def\@tempb {arXiv}\fi \@ifundefined
  {mn@eprint@\@tempb}{\@tempb:\@tempc}{\expandafter \expandafter \csname
  mn@eprint@\@tempb\endcsname \expandafter{\@tempc}}}

\bibitem[\protect\citeauthoryear{{Andreon}, {Serra}, {Moretti}  \&
  {Trinchieri}}{{Andreon} et~al.}{2016}]{2016A&A...585A.147A}
{Andreon} S.,  {Serra} A.~L.,  {Moretti} A.,   {Trinchieri} G.,  2016, \mn@doi
  [\aap] {10.1051/0004-6361/201527408}, \href
  {http://adsabs.harvard.edu/abs/2016A%26A...585A.147A} {585, A147}

\bibitem[\protect\citeauthoryear{{Astropy Collaboration} et~al.,}{{Astropy
  Collaboration} et~al.}{2013}]{2013A&A...558A..33A}
{Astropy Collaboration} et~al., 2013, \mn@doi [\aap]
  {10.1051/0004-6361/201322068}, \href
  {http://adsabs.harvard.edu/abs/2013A%26A...558A..33A} {558, A33}

\bibitem[\protect\citeauthoryear{{Bleem} et~al.,}{{Bleem}
  et~al.}{2015}]{2015ApJS..216...27B}
{Bleem} L.~E.,  et~al., 2015, \mn@doi [\apjs] {10.1088/0067-0049/216/2/27},
  \href {http://adsabs.harvard.edu/abs/2015ApJS..216...27B} {216, 27}

\bibitem[\protect\citeauthoryear{{B{\"o}hringer} et~al.,}{{B{\"o}hringer}
  et~al.}{2004}]{2004A&A...425..367B}
{B{\"o}hringer} H.,  et~al., 2004, \mn@doi [\aap] {10.1051/0004-6361:20034484},
  \href {http://adsabs.harvard.edu/abs/2004A%26A...425..367B} {425, 367}

\bibitem[\protect\citeauthoryear{{B{\"o}hringer} et~al.,}{{B{\"o}hringer}
  et~al.}{2010}]{2010A&A...514A..32B}
{B{\"o}hringer} H.,  et~al., 2010, \mn@doi [\aap]
  {10.1051/0004-6361/200913911}, \href
  {http://adsabs.harvard.edu/abs/2010A%26A...514A..32B} {514, A32}

\bibitem[\protect\citeauthoryear{{B{\"o}hringer}, {Chon}  \&
  {Collins}}{{B{\"o}hringer} et~al.}{2014}]{2014A&A...570A..31B}
{B{\"o}hringer} H.,  {Chon} G.,   {Collins} C.~A.,  2014, \mn@doi [\aap]
  {10.1051/0004-6361/201323155}, \href
  {http://adsabs.harvard.edu/abs/2014A%26A...570A..31B} {570, A31}

\bibitem[\protect\citeauthoryear{{Bryan} \& {Norman}}{{Bryan} \&
  {Norman}}{1998}]{1998ApJ...495...80B}
{Bryan} G.~L.,  {Norman} M.~L.,  1998, \mn@doi [\apj] {10.1086/305262}, \href
  {http://adsabs.harvard.edu/abs/1998ApJ...495...80B} {495, 80}

\bibitem[\protect\citeauthoryear{{Chon} \& {B{\"o}hringer}}{{Chon} \&
  {B{\"o}hringer}}{2017}]{2017A&A...606L...4C}
{Chon} G.,  {B{\"o}hringer} H.,  2017, \mn@doi [\aap]
  {10.1051/0004-6361/201731854}, \href
  {http://adsabs.harvard.edu/abs/2017A%26A...606L...4C} {606, L4}

\bibitem[\protect\citeauthoryear{{Chon}, {B{\"o}hringer}  \& {Smith}}{{Chon}
  et~al.}{2012}]{2012A&A...548A..59C}
{Chon} G.,  {B{\"o}hringer} H.,   {Smith} G.~P.,  2012, \mn@doi [\aap]
  {10.1051/0004-6361/201220267}, \href
  {http://adsabs.harvard.edu/abs/2012A%26A...548A..59C} {548, A59}

\bibitem[\protect\citeauthoryear{{De Boni}, {Ettori}, {Dolag}  \&
  {Moscardini}}{{De Boni} et~al.}{2013}]{2013MNRAS.428.2921D}
{De Boni} C.,  {Ettori} S.,  {Dolag} K.,   {Moscardini} L.,  2013, \mn@doi
  [\mnras] {10.1093/mnras/sts235}, \href
  {http://adsabs.harvard.edu/abs/2013MNRAS.428.2921D} {428, 2921}

\bibitem[\protect\citeauthoryear{{Diemer}, {More}  \& {Kravtsov}}{{Diemer}
  et~al.}{2013}]{2013ApJ...766...25D}
{Diemer} B.,  {More} S.,   {Kravtsov} A.~V.,  2013, \mn@doi [\apj]
  {10.1088/0004-637X/766/1/25}, \href
  {http://adsabs.harvard.edu/abs/2013ApJ...766...25D} {766, 25}

\bibitem[\protect\citeauthoryear{{Dolag} \& {Stasyszyn}}{{Dolag} \&
  {Stasyszyn}}{2009}]{2009MNRAS.398.1678D}
{Dolag} K.,  {Stasyszyn} F.,  2009, \mn@doi [\mnras]
  {10.1111/j.1365-2966.2009.15181.x}, \href
  {http://adsabs.harvard.edu/abs/2009MNRAS.398.1678D} {398, 1678}

\bibitem[\protect\citeauthoryear{{Dolag}, {Jubelgas}, {Springel}, {Borgani}  \&
  {Rasia}}{{Dolag} et~al.}{2004}]{2004ApJ...606L..97D}
{Dolag} K.,  {Jubelgas} M.,  {Springel} V.,  {Borgani} S.,   {Rasia} E.,  2004,
  \mn@doi [\apjl] {10.1086/420966}, \href
  {http://adsabs.harvard.edu/abs/2004ApJ...606L..97D} {606, L97}

\bibitem[\protect\citeauthoryear{{Duffy}, {Schaye}, {Kay}  \& {Dalla
  Vecchia}}{{Duffy} et~al.}{2008}]{2008MNRAS.390L..64D}
{Duffy} A.~R.,  {Schaye} J.,  {Kay} S.~T.,   {Dalla Vecchia} C.,  2008, \mn@doi
  [\mnras] {10.1111/j.1745-3933.2008.00537.x}, \href
  {http://adsabs.harvard.edu/abs/2008MNRAS.390L..64D} {390, L64}

\bibitem[\protect\citeauthoryear{{Fabjan}, {Borgani}, {Tornatore}, {Saro},
  {Murante}  \& {Dolag}}{{Fabjan} et~al.}{2010}]{2010MNRAS.401.1670F}
{Fabjan} D.,  {Borgani} S.,  {Tornatore} L.,  {Saro} A.,  {Murante} G.,
  {Dolag} K.,  2010, \mn@doi [\mnras] {10.1111/j.1365-2966.2009.15794.x}, \href
  {http://adsabs.harvard.edu/abs/2010MNRAS.401.1670F} {401, 1670}

\bibitem[\protect\citeauthoryear{{Hasselfield} et~al.,}{{Hasselfield}
  et~al.}{2013}]{2013JCAP...07..008H}
{Hasselfield} M.,  et~al., 2013, \mn@doi [\jcap]
  {10.1088/1475-7516/2013/07/008}, \href
  {http://adsabs.harvard.edu/abs/2013JCAP...07..008H} {7, 008}

\bibitem[\protect\citeauthoryear{{Hirschmann}, {Dolag}, {Saro}, {Bachmann},
  {Borgani}  \& {Burkert}}{{Hirschmann} et~al.}{2014}]{2014MNRAS.442.2304H}
{Hirschmann} M.,  {Dolag} K.,  {Saro} A.,  {Bachmann} L.,  {Borgani} S.,
  {Burkert} A.,  2014, \mn@doi [\mnras] {10.1093/mnras/stu1023}, \href
  {http://adsabs.harvard.edu/abs/2014MNRAS.442.2304H} {442, 2304}

\bibitem[\protect\citeauthoryear{Hunter}{Hunter}{2007}]{Hunter:2007}
Hunter J.~D.,  2007, \mn@doi [Computing In Science \& Engineering]
  {10.1109/MCSE.2007.55}, 9, 90

\bibitem[\protect\citeauthoryear{{Komatsu} et~al.,}{{Komatsu}
  et~al.}{2011}]{2011ApJS..192...18K}
{Komatsu} E.,  et~al., 2011, \mn@doi [\apjs] {10.1088/0067-0049/192/2/18},
  \href {http://adsabs.harvard.edu/abs/2011ApJS..192...18K} {192, 18}

\bibitem[\protect\citeauthoryear{{Ludlow}, {Bose}, {Angulo}, {Wang},
  {Hellwing}, {Navarro}, {Cole}  \& {Frenk}}{{Ludlow}
  et~al.}{2016}]{2016MNRAS.460.1214L}
{Ludlow} A.~D.,  {Bose} S.,  {Angulo} R.~E.,  {Wang} L.,  {Hellwing} W.~A.,
  {Navarro} J.~F.,  {Cole} S.,   {Frenk} C.~S.,  2016, \mn@doi [\mnras]
  {10.1093/mnras/stw1046}, \href
  {http://adsabs.harvard.edu/abs/2016MNRAS.460.1214L} {460, 1214}

\bibitem[\protect\citeauthoryear{{Macci{\`o}}, {Dutton}  \& {van den
  Bosch}}{{Macci{\`o}} et~al.}{2008}]{2008MNRAS.391.1940M}
{Macci{\`o}} A.~V.,  {Dutton} A.~A.,   {van den Bosch} F.~C.,  2008, \mn@doi
  [\mnras] {10.1111/j.1365-2966.2008.14029.x}, \href
  {http://adsabs.harvard.edu/abs/2008MNRAS.391.1940M} {391, 1940}

\bibitem[\protect\citeauthoryear{{Mantz} et~al.,}{{Mantz}
  et~al.}{2016}]{2016MNRAS.463.3582M}
{Mantz} A.~B.,  et~al., 2016, \mn@doi [\mnras] {10.1093/mnras/stw2250}, \href
  {http://adsabs.harvard.edu/abs/2016MNRAS.463.3582M} {463, 3582}

\bibitem[\protect\citeauthoryear{{Neto} et~al.}{{Neto}
  et~al.}{2007}]{2007MNRAS.381.1450N}
{Neto} A.~F.,  et~al., 2007, \mn@doi [\mnras]
  {10.1111/j.1365-2966.2007.12381.x}, \href
  {http://adsabs.harvard.edu/abs/2007MNRAS.381.1450N} {381, 1450}

\bibitem[\protect\citeauthoryear{{Planck Collaboration} et~al.,}{{Planck
  Collaboration} et~al.}{2011}]{2011A&A...536A...8P}
{Planck Collaboration} et~al., 2011, \mn@doi [\aap]
  {10.1051/0004-6361/201116459}, \href
  {http://adsabs.harvard.edu/abs/2011A%26A...536A...8P} {536, A8}

\bibitem[\protect\citeauthoryear{{Planck Collaboration} et~al.,}{{Planck
  Collaboration} et~al.}{2016a}]{2016A&A...594A..24P}
{Planck Collaboration} et~al., 2016a, \mn@doi [\aap]
  {10.1051/0004-6361/201525833}, \href
  {http://adsabs.harvard.edu/abs/2016A%26A...594A..24P} {594, A24}

\bibitem[\protect\citeauthoryear{{Planck Collaboration} et~al.,}{{Planck
  Collaboration} et~al.}{2016b}]{2016A&A...594A..27P}
{Planck Collaboration} et~al., 2016b, \mn@doi [\aap]
  {10.1051/0004-6361/201525823}, \href
  {http://adsabs.harvard.edu/abs/2016A%26A...594A..27P} {594, A27}

\bibitem[\protect\citeauthoryear{{Pratt}, {Croston}, {Arnaud}  \&
  {B{\"o}hringer}}{{Pratt} et~al.}{2009}]{2009A&A...498..361P}
{Pratt} G.~W.,  {Croston} J.~H.,  {Arnaud} M.,   {B{\"o}hringer} H.,  2009,
  \mn@doi [\aap] {10.1051/0004-6361/200810994}, \href
  {http://adsabs.harvard.edu/abs/2009A%26A...498..361P} {498, 361}

\bibitem[\protect\citeauthoryear{{Rossetti}, {Gastaldello}, {Eckert}, {Della
  Torre}, {Pantiri}, {Cazzoletti}  \& {Molendi}}{{Rossetti}
  et~al.}{2017}]{2017MNRAS.468.1917R}
{Rossetti} M.,  {Gastaldello} F.,  {Eckert} D.,  {Della Torre} M.,  {Pantiri}
  G.,  {Cazzoletti} P.,   {Molendi} S.,  2017, \mn@doi [\mnras]
  {10.1093/mnras/stx493}, \href
  {http://adsabs.harvard.edu/abs/2017MNRAS.468.1917R} {468, 1917}

\bibitem[\protect\citeauthoryear{{Rykoff} et~al.,}{{Rykoff}
  et~al.}{2014}]{2014ApJ...785..104R}
{Rykoff} E.~S.,  et~al., 2014, \mn@doi [\apj] {10.1088/0004-637X/785/2/104},
  \href {http://adsabs.harvard.edu/abs/2014ApJ...785..104R} {785, 104}

\bibitem[\protect\citeauthoryear{{Rykoff} et~al.,}{{Rykoff}
  et~al.}{2016}]{2016ApJS..224....1R}
{Rykoff} E.~S.,  et~al., 2016, \mn@doi [\apjs] {10.3847/0067-0049/224/1/1},
  \href {http://adsabs.harvard.edu/abs/2016ApJS..224....1R} {224, 1}

\bibitem[\protect\citeauthoryear{{Saro} et~al.,}{{Saro}
  et~al.}{2017}]{2017MNRAS.468.3347S}
{Saro} A.,  et~al., 2017, \mn@doi [\mnras] {10.1093/mnras/stx594}, \href
  {http://adsabs.harvard.edu/abs/2017MNRAS.468.3347S} {468, 3347}

\bibitem[\protect\citeauthoryear{{Springel}}{{Springel}}{2005}]{2005MNRAS.364.1105S}
{Springel} V.,  2005, \mn@doi [\mnras] {10.1111/j.1365-2966.2005.09655.x},
  \href {http://adsabs.harvard.edu/abs/2005MNRAS.364.1105S} {364, 1105}

\bibitem[\protect\citeauthoryear{{Springel} \& {Hernquist}}{{Springel} \&
  {Hernquist}}{2003}]{2003MNRAS.339..289S}
{Springel} V.,  {Hernquist} L.,  2003, \mn@doi [\mnras]
  {10.1046/j.1365-8711.2003.06206.x}, \href
  {http://adsabs.harvard.edu/abs/2003MNRAS.339..289S} {339, 289}

\bibitem[\protect\citeauthoryear{{Springel}, {White}, {Tormen}  \&
  {Kauffmann}}{{Springel} et~al.}{2001}]{2001MNRAS.328..726S}
{Springel} V.,  {White} S.~D.~M.,  {Tormen} G.,   {Kauffmann} G.,  2001,
  \mn@doi [\mnras] {10.1046/j.1365-8711.2001.04912.x}, \href
  {http://adsabs.harvard.edu/abs/2001MNRAS.328..726S} {328, 726}

\bibitem[\protect\citeauthoryear{{Springel}, {Di Matteo}  \&
  {Hernquist}}{{Springel} et~al.}{2005}]{2005MNRAS.361..776S}
{Springel} V.,  {Di Matteo} T.,   {Hernquist} L.,  2005, \mn@doi [\mnras]
  {10.1111/j.1365-2966.2005.09238.x}, \href
  {http://adsabs.harvard.edu/abs/2005MNRAS.361..776S} {361, 776}

\bibitem[\protect\citeauthoryear{{Thompson}}{{Thompson}}{2014}]{2014ascl.soft11001T}
{Thompson} R.,  2014, {pyGadgetReader: GADGET snapshot reader for python},
  Astrophysics Source Code Library (\mn@eprint {ascl} {1411.001})

\bibitem[\protect\citeauthoryear{{Tornatore}, {Borgani}, {Springel},
  {Matteucci}, {Menci}  \& {Murante}}{{Tornatore}
  et~al.}{2003}]{2003MNRAS.342.1025T}
{Tornatore} L.,  {Borgani} S.,  {Springel} V.,  {Matteucci} F.,  {Menci} N.,
  {Murante} G.,  2003, \mn@doi [\mnras] {10.1046/j.1365-8711.2003.06631.x},
  \href {http://adsabs.harvard.edu/abs/2003MNRAS.342.1025T} {342, 1025}

\bibitem[\protect\citeauthoryear{{Tornatore}, {Borgani}, {Dolag}  \&
  {Matteucci}}{{Tornatore} et~al.}{2007}]{2007MNRAS.382.1050T}
{Tornatore} L.,  {Borgani} S.,  {Dolag} K.,   {Matteucci} F.,  2007, \mn@doi
  [\mnras] {10.1111/j.1365-2966.2007.12070.x}, \href
  {http://adsabs.harvard.edu/abs/2007MNRAS.382.1050T} {382, 1050}

\makeatother
\end{thebibliography}



\appendix 

\section{Substructure mass fraction outside $R_{\texorpdfstring{\MakeLowercase{\mathrm{vir}}}{\mathrm{vir}}}$} \label{outskirts}

In this appendix, we study the global distribution of substructures in clusters as defined by SUBFIND, with neither any radial limit nor any mass threshold. This is a fully consistent internal SUBFIND analysis, completely independent of the background density of the Universe and of any overdensity definition. For every cluster in our sample, we define the quantity $f_{\mathrm{sub,global}}$ as the ratio between the sum of the masses of all satellite halos detected by SUBFIND and the mass of the main halo, $M_{\mathrm{mh}}$. The results are reported in Table \ref{f_sub_out_table}. As for $f_{\mathrm{sub,vir}}$, also for $f_{\mathrm{sub}}$ there is only a weak correlation with mass. A small number of clusters with a very high mass fraction in substructures is driving the mean values, and the difference between the mean and the median is higher than for $f_{\mathrm{sub,vir}}$.
The last redshift bin is peculiar, not only because the average value of $f_{\mathrm{sub}}$ is low compared to the other bins, but also because the scatter is low and the more extreme object has a value of $f_{\mathrm{sub}}$ that is less than half that value of the most extreme case at any other redshift.

\begin{table}
\begin{center}
\caption{Same as Table \ref{f_sub_table}, but for $f_{\mathrm{sub,global}}$.}  \label{f_sub_out_table}
\begin{tabular}{cccccc} 
\\
\hline
\hline
Redshift & mean & $\sigma$ & median & $16 \%$ & $84 \%$ \\
\hline
\hline
$z=0$ & $0.125$ & $0.119$ & $0.088$ & $0.050$ & $0.189$ \\
\hline
$z=0.137$ & $0.123$ & $0.129$ & $0.081$ & $0.051$ & $0.175$ \\
\hline
$z=0.293$ & $0.127$ & $0.112$ & $0.091$ & $0.051$ & $0.207$ \\
\hline
$z=0.470$ & $0.131$ & $0.139$ & $0.085$ & $0.053$ & $0.209$ \\
\hline
$z=0.783$ & $0.131$ & $0.163$ & $0.087$ & $0.048$ & $0.181$ \\
\hline
$z=1.179$ & $0.095$ & $0.065$ & $0.075$ & $0.045$ & $0.148$  \\
\hline
\hline
\end{tabular}
\end{center}
\end{table}

\begin{figure}
\hbox{
  \epsfig{figure=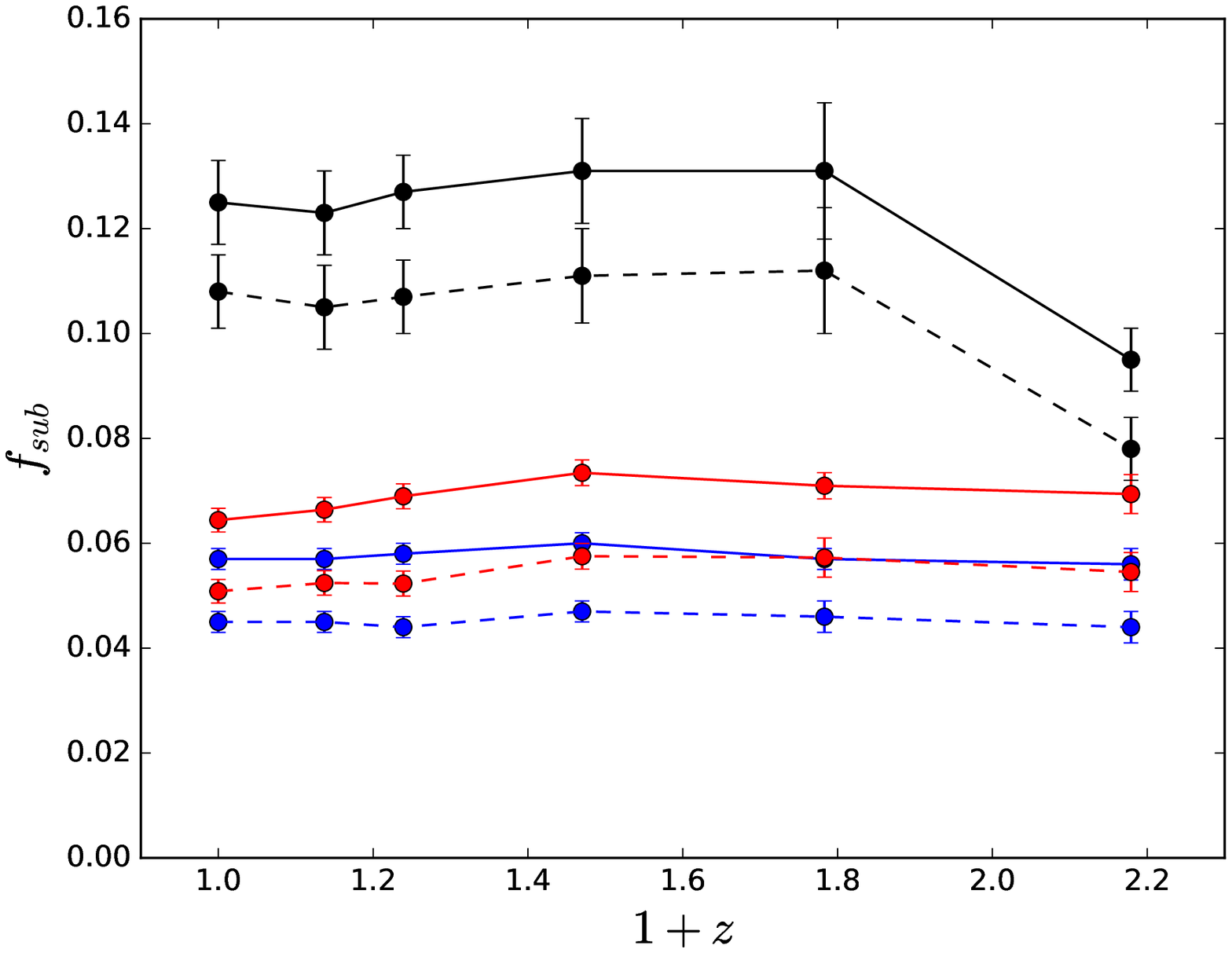,width=0.5\textwidth}
}
\caption{Evolution of the substructure mass fraction with redshift. The solid black line indicates $f_{\mathrm{sub}}$, the dashed black line indicates $f_{\mathrm{sub},>1e12}$, the solid blue line indicates $f_{\mathrm{sub,vir}}$, the dashed blue line indicates $f_{\mathrm{sub,vir},>1e12}$. Red lines indicate the same quantities as blue lines, but corrected for the mean halo-by-halo ratio $M_{\mathrm{vir}}/M_{\mathrm{mh}}$.}
\label{fsub_redshift_evolution}
\end{figure}

\noindent In Figure \ref{fsub_redshift_evolution} we compile several results on substructures as found by SUBFIND, in order to compare $f_{\mathrm{sub,global}}$ and $f_{\mathrm{sub,vir}}$. Before comparing the results, we must consider the fact that the first quantity is normalized with respect to the mass of the main halo while the latter uses the virial mass as normalization. Since the ratio $M_{\mathrm{vir}}/M_{\mathrm{mh}}$ increases with redshift, we correct $f_{\mathrm{sub,vir}}$ and $f_{\mathrm{sub,vir},>1e12}$ to take this into account. The corrected values can be seen in red in Figure \ref{fsub_redshift_evolution}. Even after the correction, it is evident that half of the mass in substructures resides outside the virial radius. And this is true also if we limit our analysis to objects more massive than $10^{12} \ {\rm M_{\odot}} \ h^{-1}$. Moreover, within the virial radius, also the results of the redshift bin $z>1$ is correlated with the others. So, the origin of the problem resides in the subhalo distribution outside the virial radius at $z>1$ which is lower than for the other redshifts.

\bsp	
\label{lastpage}
\end{document}